\documentclass[preprint]{elsarticle}
\usepackage{amsmath}
\usepackage{amsfonts}
\usepackage{amssymb}
\usepackage{amsthm}
\usepackage{graphicx}
\usepackage{hyperref}
\usepackage{color}
\usepackage{flafter}
\usepackage[normalem]{ulem}

\newcommand{\Fig}[1]{Fig.~\ref{#1}}

\newcommand{\Eq}[1]{equation~(\ref{#1})}
\newcommand{\nn}{\nonumber\\}
\newcommand{\ie}{{\emph{i.e.~}}}

\newcommand{\<}{\langle}
\renewcommand{\>}{\rangle}
\newcommand{\ket}[1]{|{#1}\rangle}
\newcommand{\bra}[1]{\langle{#1}|}
\newcommand{\be}{\begin{eqnarray}}
\newcommand{\ee}{\end{eqnarray}}
\newcommand{\bpm}{\begin{pmatrix}}
\newcommand{\epm}{\end{pmatrix}}

\newcommand{\p}{\partial}

\newcommand{\ra}{\rightarrow}
\newcommand{\lra}{\leftrightarrow}
\renewcommand{\v}[1]{{\boldsymbol{#1}}}
\renewcommand{\a}{\alpha}
\renewcommand{\b}{\beta}

\newcommand{\e}{\epsilon}
\newcommand{\s}{\sigma}
\renewcommand{\t}{\tau}
\newcommand{\g}{\gamma}
\newcommand{\G}{\Gamma}

\begin{document}

\title{A Holographic Theory for the Phase Transitions
Between Fermionic Symmetry-protected Topological States}

\author[ucb]{Lokman Tsui}
\ead{lokman@berkeley.edu}
\author[ucb]{Yen-Ta Huang}
\ead{yenta.huang@berkeley.edu}
\author[ucb,lbl]{Dung-Hai Lee\corref{cor1}}
\ead{dunghai@berkeley.edu}
\cortext[cor1]{Corresponding author}
\address[ucb]{Department of Physics, University of California, Berkeley, California 94720, USA}
\address[lbl]{Materials Sciences Division, Lawrence Berkeley National Laboratories, Berkeley, California 94720, USA}

\setlength{\parindent}{0pt}

\date{\today}

\begin{abstract}
In an earlier work\cite{Tsui2015} we developed a holographic theory for the phase transition between bosonic symmetry-protected topological (SPT) states. This paper is a continuation of it.   Here we present the holographic theory for fermionic SPT phase transitions. We show that in any dimension $ d $, the critical states of fermionic SPT phase transitions has an emergent $Z_2^T$ symmetry and can be realized on the boundary of a $ d+1 $-dimensional bulk SPT with an extra $Z_2^T$ symmetry.
\end{abstract}


\maketitle

\section{Introduction} \label{sec:intro}

Symmetry protected topological states (SPTs) are new quantum phases of matter. They are characterized by a fully gapped bulk but gapless boundary. Moreover,  as long as the symmetry of these phases are unbroken, the gapless boundary excitations survive any perturbation. (For simplicity we assume no topological order develops at the boundary.)\\ 

SPTs fall into two broad classes: bosonic and fermionic SPTs.  The Hamiltonian of bosonic SPTs consists of commuting local degrees of freedom. An example is  the Haldane phase of the spin-1 chain. In contrast, the Hamiltonian of fermionic SPTs consists of anti-commuting local degrees of freedom. They describe insulating or superconducting states of fermions.\\

For a fixed Hamiltonian symmetry, the bosonic and fermionic SPTs are classified into equivalent classes. The transition between different classes requires a quantum phase transition, accompanied by the closing of the bulk energy gap. Unlike usual 
phase transitions, these phase transitions do not involve any symmetry change. Instead, what differentiate the two phases are the conformal field theories of their boundaries. Currently the classification theories of the SPTs phases are well-developed. However, the theory describing the SPT phase transitions is still in its infancy. \\

In Ref.\cite{Tsui2015} a holographic theory for the phase transitions between a wide class of bosonic SPTs is developed. In a nutshell, it is shown that the critical state of such a phase transition can be described as the boundary state of a SPT living in one dimension higher. In addition, the higher dimensional SPT has an extra anti-unitary ($Z_2^T$) symmetry. 
The implication of this theory are (1) The excitations of the critical theory are the fluctuating boundaries between the two SPT phases. (2) The anti-unitary group $Z_2^T$ acts as an emergent duality symmetry at the SPT phase transition. In this paper we emphasize another implication of the holographic correspondence, namely, (3) in the presence of (emergent) Lorentz symmetry, the conformal spectrum of the critical theory at the SPT phase transition is the same as the ground state entanglement spectrum of the holographic bulk SPT. This in turn implies the topological classification of the bulk SPT also classifies the conformal field theory for the SPT phase transition. In this paper we also answer the important question, namely, whether there is an analogous holographic description for the fermionic SPT transitions. We develop such a theory for free and a specific type of interacting fermion SPT transitions. \\

The outline of the paper is as follows. In section \ref{freefermionclass}, we briefly review the classification results for free fermion SPT.  In section \ref{holo} we present the holographic theory for free fermionic SPT phase transitions. In section \ref{examples} we show a 1D and a 2D example of the holographic correspondence established in section \ref{holo}. In section \ref{ddw} we consider a specific interacting version of the holographic bulk SPT under the proviso that the interaction term does not collapse the bulk gap. We show that such interacting bulk SPT can be viewed as containing condensed $Z_2^T$ domain walls. In section \ref{decdw} we show that, analogous to the bosonic holographic theory\cite{Tsui2015}, each domain wall in section \ref{ddw} is decorated with a lower dimension SPT. We demonstrate this by numerics using the 1D and 2D examples. The analytic proof of the statement is given in \ref{apdx:ddw}. In section \ref{bint} we discuss the boundary of a specific kind of interacting bulk theory. In section \ref{phtrintspt} we discuss phase transitions between interacting SPT phases whose critical theory is the boundary theory in section \ref{bint}. We argue that depending on whether the $Z_2^T$ symmetry is spontaneously broken, such interacting boundary theories either describe continuous or first order SPT phase transitions. 
In section \ref{bbc} we discuss the correspondence between the entanglement spectrum of the holographic bulk SPT and the conformal spectrum at the critical point of SPT phase transition. 
\\

In addition to the main text outlined above there are ten appendices. Their contents are summarized as follows. In \ref{regr} we present the rules for regularizing a continuum field theories of SPTs on a hyper-cubic lattice. 
In  \ref{apdx:m0} we prove that the minimal models defined in the main text, which describes a pair of inequivalent SPTs, must have one allowed mass term only. Using continuum field theory, in \ref{apdx:dwgapless} we prove the existence of gapless modes at the interface between the two inequivalent SPT phases in minimal models. In \ref{reg} we relate the interface gapless modes in \ref{apdx:dwgapless} to the boundary gapless modes of regularized lattice theory of non-trivial SPTs. In \ref{classt} we summarize the topological classification for free fermion SPTs protected by the $T\rtimes Q\ltimes C$ symmetries. In the same appendix we specify the dimension of the gamma and mass matrices in the minimal models. In \ref{unique} we prove that if a SPT transition is described by a minimal model, its holographic bulk must also be described by a minimal model. In \ref{apdx:ddw} we show analytically that the holographic bulk SPT has a decorated-$ Z_2^T $ domain wall interpretation. Namely, if a domain wall is statically frozen, its associated mode space Hamiltonian can be block-diagonalized with a sub-block describing localized degrees of freedom on the domain wall. The Hamiltonian in this sub-block is  that of a non-trivial SPT. In \ref{apdx:lat} we present the lattice Hamiltonian used in the numerical studies of the $Z_2^T$ domain walls in the main text. In the last appendix, \ref{apdx:spacetimelat}, we discuss the 
space-time rotation necessary to establish the correspondence between the ground state wavefunction of the holographic bulk and the Boltzmann weight of the conformal field theory at the SPT critical point.
\\

\section{The free fermion classification} \label{freefermionclass}
\subsection{The low energy effective Hamiltonian}
Free fermionic SPT may be classified\cite{Schnyder2008,Kitaev2009,Wen2012} by looking at their low energy effective Hamiltonians. These Hamiltonians have the following form
\begin{align}
H=\int d^d x~ X^T(\v x) \left[\sum_{j=1}^{D}-i\Gamma_j\p_j +i \lambda M\right]X(\v x).\label{eq:Hf}
\end{align}
Here $ X(\v x) $ is a $ n$-component Majorana field operator. We use the Majorana fermion representation so that it can describe the Bogoliubov excitations of a superconductor. Different components of $X$ are labeled by the spin, orbital and the Majorana indices. The Majorana index labels the real and imaginary part of a complex fermion operator. If $X$ has $n$ components, the matrices $\G_j$ and $M$ in \Eq{eq:Hf} are all $n\times n$ matrices. In the following we shall refer to the $n$ dimensional internal space of the Majorana fermion as the ``mode space''. 
The real-valued symmetric matrices $ \Gamma_i $ obey the Clifford algebra $ \{\Gamma_j,\Gamma_k\}=2\delta_{jk} I_{n} $, where $ I_{n} $ is the $ n\times n $ identity matrix. $ M $ is an $ n\times n $ antisymmetric real matrix satisfying $ \{\Gamma_j,M\}=0 $ for $ j=1,...,n $. It causes the energy gap. We require  $ M^2=-I_{n} $ so that the absolute value of $\lambda$ (a real parameter) sets the size of the energy gap.
\\
\subsection{The symmetries}
In this paper we focus on on-site symmetries, i.e., symmetries that acts on the degrees of freedom on each lattice site independently.  Let $G$ be such a symmetry group.The action of an element $\hat{g}\in G $ on the Majorana fermion obeys $$ \hat{g}X\hat{g}^{-1}=g X. $$ where $g$ is an $n\times n$ orthogonal matrix since it preserves the anti-commutation relations between Majorana operators $ \{(gX)_i,(gX)_j\} = \{X_i,X_j\} = 2 \delta_{ij} \hat{I} $. If $ \hat{g} $ is unitary, the associated $ g $ commutes with $ \Gamma_i $ and $ M $. If $ \hat{g} $ is anti-unitary, it negates the $ i $ in front of the kinetic and the mass term, thus requiring $ g $ to anti-commute with $ \Gamma_i $ and $ M $.\\

\subsection{The minimal model}
Given a symmetry group $ G $, we define $ n_0 $ as the minimum value of $ n $ for which mass term(s) $ M $ satisfying all the above requirements exist. The corresponding model given in  \Eq{eq:Hf} is called the ``minimal model''. The minimal models are the ``atoms'' in the SPT world. A minimal model can describe either  (a) a trivial SPT or, (b) a pair of inequivalent SPTs. In the following we focus on the more interesting case, namely case (b).\\

\begin{figure}[h]
\centering
\includegraphics[scale=1]{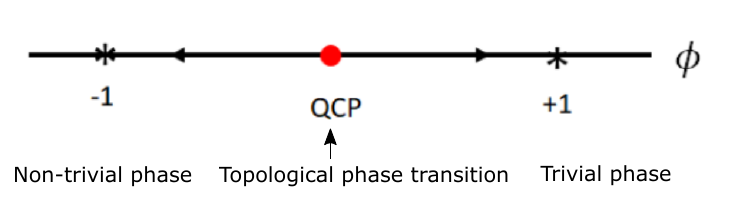}
\caption{(Color online) Schematic phase diagram of \eqref{eq:Hf1}. For $ \phi<0 $ ($ \phi>0 $) the Hamiltonian describes a (non-)trivial SPT phase. Phase transition occurs at $ \phi=0 $ (depicted by red dot).}
\label{fig:QCP}
\end{figure}

For case (b) it can be shown that when $ n=n_0 $ there is only one ($n_0\times n_0$) mass matrix $m_0$ that satisfies $ \{m_0,\g_i\}=0 $  and is symmetric under $ G $ (see \ref{apdx:m0} for a proof). 
We denote the Hamiltonian of such a minimal model by
\begin{align}
H=\int d^d x~ \chi^T(\v x) \left[\sum_{j=1}^{d}-i\g_j\p_j +i \phi~m_0\right]\chi(\v x).\label{eq:Hf1}
\end{align}
Note that  we have switched to the lower case symbols. This is  to emphasize it is a minimal model. In \Eq{eq:Hf1} the real scalar $ \phi$ is the mass parameter and  $ \phi>0 $ and $ \phi<0 $ corresponds to the two inequivalent SPT phases.
The fact that \Eq{eq:Hf1} with opposite sign of $\phi$ describes inequivalent SPTs can be shown by considering a domain wall separating the spatial regions with $\phi>0$ and $\phi<0$. In \ref{apdx:dwgapless} we show the existence of gapless fermion modes localizing on the wall. 
Such gapless fermion modes signify the topological inequivalence of the SPTs. By tuning $ \phi$ to zero, the fermion gap vanishes hence it marks the phase transition between the two SPT phases (see \Fig{fig:QCP}).\\ 
\\

\subsection{The regularized topological non-trivial minimal models}
So far we have been discussing continuum field theories. A non-trivial lattice SPT Hamiltonian is a regularized Hamiltonian which reduces to \Eq{eq:Hf1} in the low energy limit. When such a lattice Hamiltonian is subjected to the open boundary condition, it yields gapless boundary modes. In \ref{regr} we give the rules for obtaining such a regularized Hamiltonian on the hypercubic lattice.
Here we simply summarize the results.  Upon Fourier transformation \Eq{eq:Hf1} becomes
\be
H=\sum_{\v k\in B(0)}\chi^T(-\v k) \left[\sum_{j=1}^{d} k_j\g_j+i \phi~m_0\right]\chi(\v k).\label{eq:Hf1k}
\ee
Here $B(0)$ denotes a small ball around $\v k=0$, and $\chi(\v k)$ is the Fourier transform of the Majorana field $\chi(\v x)$. The regularized lattice Hamiltonian corresponding to \Eq{eq:Hf1k} read
\be
H=\sum_{\v k\in {\rm BZ}}\chi^T(-\v k) \left[\sum_{j=1}^{d} \sin k_j\g_j+i (d-\sum_{j=1}^{d} \cos k_j)m_0+i ~\phi~m_0\right]\chi(\v k).\label{hk2}
\ee
Here ``BZ'' stands for the Brillouin zone of the $ d $-dimensional  hypercubic lattice. (See \ref{regr} for how to obtain the real space version of \Eq{hk2}.) When $\phi\ne 0$ the second term removes the unwanted  gapless nodes at all time-reversal invariant $\v k$ points except $\v k=0$. (A time-reversal invariant $\v k$ point satisfies $-\v k=\v k+\v G$ where $\v G$ is a reciprocal lattice vector.) Equation \ref{hk2} describes a non-trivial SPT when $\phi<0$. In \ref{reg} we prove that the boundary gapless modes of \Eq{hk2} with $ \phi<0 $ is the same as those at the interface between two regions described by \Eq{eq:Hf1} with opposite $\phi$.  \\

\subsection{Stacking the minimal models}
\begin{figure}[h]
\centering
\includegraphics[scale=0.6]{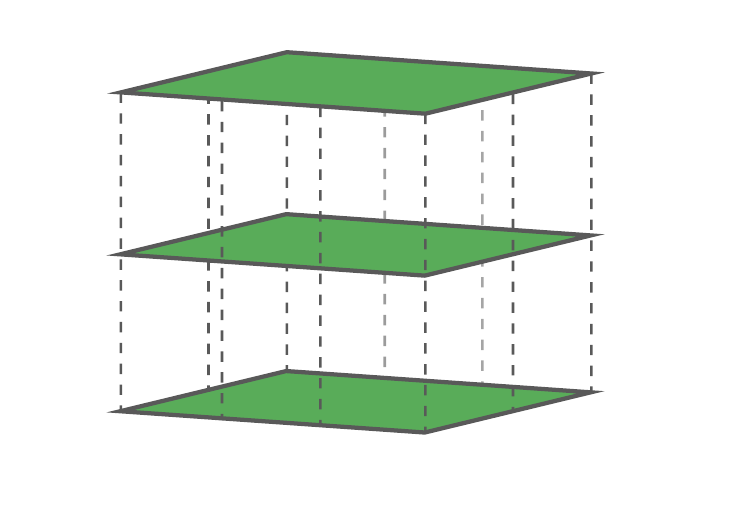}
\caption{(Color online) A general SPT can be regarded as the stacking of layers of minimal SPT's(depicted in green), and turning on symmetry-respecting interactions across the layers(depicted by dotted lines). }
\label{fig:stacking}
\end{figure}

The Hamiltonian describing general non-trivial SPT phases are constructed by 
``stacking'' together the non-trivial minimum lattice models (\Fig{fig:stacking}).
Stacking can be achieved by taking a direct sum of the mode space, 
and turning on any symmetry-allowed interaction (by ``interaction" here we mean two-fermion operators, acting across different layers, not to be confused by the ``interacting systems" considered in sections \ref{ddw} to \ref{bbc}, where it means the addition of four-fermion(or higher order) operators.) between the degrees of freedom associated with the minimal models. Sometimes stacking can produce an infinite number of different topological phases. In which case the SPT in question is $ Z $ classified. Other times stacking produces at most two different phases. In this case the SPT in question is $ Z_2 $ classified. In \ref{classt} we summarize the classification result of free fermions with on-site $T\rtimes Q\ltimes C$ symmetries. The values of $n_0$ for the minimal models are also given.\\

It can be shown that for any pair of non-minimal SPTs, the phase transition between the two can be deformed into subsequent phase transitions between minimal model SPTs. For the same reason understanding the phase transition between SPTs described by the minimal model constitutes a complete understanding of the SPT phase transitions. 

\section{The holographic theory}\label{holo}
We begin this section by asking ``is there a  symmetry group which can protect the $\phi=0$ critical point of \Eq{eq:Hf1k} or \Eq{hk2}.'' The answer is yes. The symmetry group can be constructed by adding the generator of a two-element anti-unitary group, $Z_2^T$, to $G$. Specifically, such generator sends $\chi$ to  $m_0\chi$ (recall that $m_0$ is an $n_0\times n_0$ matrix). Because $m_0$ anti-commutes with all $\g_j$'s, it fulfills the requirement of being the representation of an anti-unitary symmetry. Moreover, this additional anti-unitary symmetry changes the sign of the only allowed mass term ($ i~\phi~m_0$)(the sign reversal is caused by the complex conjugation) hence forbids it. In the remaining of the paper we shall denote the generator of the $Z_2^T$ by $T$. However it is important to remember that this generator does not necessarily correspond to the usual time reversal transformation. With the extra $Z_2^T$ symmetry the resulting enlarged symmetry group protects the gapless critical state described by \Eq{eq:Hf1}(or \Eq{eq:Hf1k}) at $\phi=0$. Note that we talk about the continuum field theory rather than the lattice model in \Eq{hk2} at $\phi=0$. This is because the regularization term in \Eq{hk2} breaks the $Z_2^T$ symmetry. (Therefore the $ Z_2^T $ is an emergent symmetry at the boundary dimension.) The above arguments suggest the possible existence of a  SPT  in one higher dimension which is protected by this enlarged symmetry group, and has the $\phi=0$ critical theory as its boundary theory. The reader might wonder that on the boundary theory which does not have regularization, whether non-trivial or trivial SPT have absolute meaning. Here we emphasize that they still have relative meaning, in the sense that the interface where $ \phi $ changes sign carries gapless edge mode. The critical theory, which is low energy property of the theory, is captured by the boundary theory. For the sake of easy reference, in the rest of the paper  we shall denote the enlarged symmetry group as $G\times Z_2^T$. However, this notation is not meant to imply that the generator of the $Z_2^T$ commutes with the original generators of $G$.\\   

In the following we explicitly construct the Hamiltonian for the $ d+1 $- dimensional SPT. This Hamiltonian must reduce to \Eq{eq:Hf1k} (with $\phi=0$) at its boundary.  Moreover, the $Z_2^T$ symmetry must acts on the boundary fermion modes as $\chi\ra m_0\chi$. The continuum field theory for such a $d+1$ dimensional Hamiltonian is given by 
\be
H=\sum_{\v k\in B(0)}~ X^T(-\v k) \left[\sum_{j=1}^{d+1}k_j\Gamma_j+i \lambda M\right]X(\v k),\label{hbk}\ee
or its real space version
\be H=\int d^{d+1}x  X^T(\v x) \left[\sum_{j=1}^{d+1}-i\Gamma_j\p_j+i \lambda M\right]X(\v x). \label{hbx}
\ee
Here the matrix dimension of $\Gamma_j$ and $M$ are twice of that of $\g_j$ and $m_0$, i.e., $2n_0\times 2n_0$. 
As discussed in \ref{regr} the lattice regularized  version of \Eq{hbk} is given by
\be
H=\sum_{\v k\in {\rm BZ}}X^T(-\v k) \left[\sum_{j=1}^{d+1} \sin k_j\G_j+i (D-\sum_{j=1}^{D} \cos k_j)M+i ~\lambda~M\right]X(\v k),\label{hbk2}
\ee
and we refer the readers to \ref{regr} for the real space version.
In Table \ref{tab:bulkbdd} we summarize the relation between the $d$ and $d+1$ dimensional Hamiltonian, and the representation of the symmetry generators in the mode space.
\begin{table}[h]
\caption{The $ d+1 $-dimensional bulk theory whose boundary describes a $ d $-dimensional SPT transition.}
\centerline{
\begin{tabular}{|c|c|c|}
\hline 
 & $ d $-dimension boundary & $ d+1 $-dimension bulk \\ 
\hline
Fermion field & $\chi(\v x)=PX(\v x)P$ & $X(\v x)$\\
\hline
Hamiltonian &\Eq{eq:Hf1k}  &\Eq{hbk} or \Eq{hbk2} \\ 
\hline 
Matrix dimension $ n $ & $ n_0 $ & $ 2n_0 $ \\
\hline 
Gamma matrices & $ \g_1,...\g_d $ & $ \G_j=\g_j\otimes \t_z $, $ \G_{d+1}=I_{n_0}\otimes \t_x $ \\ 
\hline 
Mass matrix & $m_0$ & $ M = I_{n_0}\otimes i\t_y $ \\ 
\hline 
Symmetry group &$G$ for $\phi \ne 0$ and $G\times Z_2^T$ for $\phi=0$  &$G\times Z_2^T$\\ 
\hline 
Generator of $ Z_2^T $(i.e. $T$) & $ m_0 $ & $ m_0 \otimes \t_z$ \\ 
\hline 
Unitary symmetry generators & $ u_{\a} $ & $ U_{\a}=u_{\a} \otimes \t_0 $ \\ 
\hline 
Anti-unitary symmetry generators & $ a_{\b} $ & $ A_{\b}=a_{\b}\otimes \t_z $ \\ 
\hline 
\end{tabular} 
}\label{tab:bulkbdd}
\end{table}
\\

In the first row of Table \ref{tab:bulkbdd} the projection operator $P$ acts in the mode space it projects the fermion operator into the $\G_{d+1}M=-1$ sector. In the following we briefly explain Table \ref{tab:bulkbdd} and refer the reader to \ref{reg},\ref{apdx:dwgapless} and \ref{unique} for details. In \ref{reg} we have shown that the Hamiltonian for the boundary gapless modes of \Eq{hbk2} for $ \lambda<0 $ is the same as that localized on the domain wall between the $\lambda>0$ and $\lambda<0$ phases of the continuum field theory in \Eq{hbx}. As shown in \ref{apdx:dwgapless}, the mode space of the interface is the $ \G_{d+1}M=-1 $ subspace of the bulk theory. Since $ \G_{d+1}M=-I_{n_0}\otimes \t_z $, this restriction requires $ \t_z $  to be $+1$. Then it is  immediate that such restriction reduces the bulk $ \G_i $ to  $ \g_i $ for $ i=1,\dots,d $, and the bulk unitary/anti-unitary symmetries would also reduce to the corresponding symmetry generators for the $ d $-dimensional theory. Note that the generator of the extra $ Z_2^T $ symmetry, when restricted to the $ \G_{d+1}M=-1 $ subspace, is represented by the matrix $ m_0 $, the same as the $ d $-dimensional mass matrix. It turns out that the bulk mass $ M $ in Table \ref{tab:bulkbdd} is the only mass term capable of opening a gap while consistent with the bulk symmetries $ U_{\a},A_{\b} $ and $ Z_2^T $. The proof is presented in \ref{unique}. This implies the bulk SPT 
constructed according to Table \ref{tab:bulkbdd} is actually a minimal model. \\

In the following we provide two examples of the application of Table \ref{tab:bulkbdd}.\\

\section{Two simple examples}\label{examples}
We derive the $Z_2^T$ symmetry and the bulk Hamiltonians for these two examples using Table \ref{tab:bulkbdd}. In the following we only present the bulk Hamiltonian in the continuum form. The lattice version of it can be obtained by following the regularization rules summarized in \ref{regr}.

\subsection{An 1D example}\label{example1d}
The first example is a 1D  topological insulator protected by charge conservation and the particle-hole symmetry $ C^2=+1 $. According to the table in \ref{classt} it has the $ Z_2 $ classification and $n_0=4 $. Due to the charge conservation the Hamiltonian can be written in terms of a 4-component Majorana fermion field, or, equivalently, a 2-component complex fermion field $ \psi $  as
\be
H_1=\int dx~\psi^{\dagger}(x)\left[-i\s_z\p_x+\phi~ m_0\right] \psi(x) \label{eq:H1d}
\ee
Here the kinetic term describes a non-chiral, helical, dispersion, and $$ m_0=\s_x. $$  

The charge $ U(1) $ symmetry transforms $\psi\ra e^{i\theta}\psi$, and the particle-hole symmetry transforms $\psi \ra C \psi^{\dagger} $, where $C=\s_z$. The two inequivalent SPT phases are associated with $ \phi>0 $ and $ \phi<0 $, respectively.\\

Note that if we fine tune $\phi$ to zero,  \Eq{eq:H1d} possesses an extra anti-unitary symmetry, namely the time reversal symmetry
$ T=i\s_y $. This symmetry requires $ \phi=0 $. The critical theory at $ \phi=0 $ is the boundary theory of the 2D topological insulator  
described by the following  Hamiltonian:
\begin{align}
H_2 = \int d^2x~ \Psi^{\dagger} (x)\left[-i\s_z \t_z \p_x -i \s_0 \t_x \p_y +\lambda \s_0 \t_y \right] \Psi(x) \label{eq:hbulk}
\end{align}
In \Eq{eq:hbulk} the last term is the mass term ($M=\s_0 \t_y $). For the 2D bulk, the time reversal and particle-hole symmetry are represented by   $ T=i\s_y\t_z $ and $ C:~\psi_B \ra \s_z\t_0 \psi_B^{\dagger} $, respectively. Thus the bulk symmetry is $ G\times Z_2^T $ where $ G = U(1)\ltimes C $. In \ref{apdx:example} we show that \Eq{eq:hbulk} is the complex fermion version of the Majorana fermion Hamiltonian derived directly from Table \ref{tab:bulkbdd}.
\\

The 2D SPTs with $ G\times Z_2^T $ symmetry has  $Z_2$ classification, and \Eq{eq:hbulk} is the complex fermion version of the minimal model. When the bulk symmetries are respected, the lattice version of \Eq{eq:hbulk} (see \ref{regr}) with $\lambda<0$ possesses gapless boundary modes (see \ref{reg}). 
The Hamiltonian for such gapless modes is exactly  \Eq{eq:H1d} at $\phi=0$, namely, the critical theory of the 1D SPT phase transition.

\subsection{A 2D example}

The second example is a 2D superconductor with no symmetry (i.e.$G=\emptyset$). According to \ref{classt} the minimal model has $n_0=2$ and the classification is $Z$. In the Majorana fermion representation
the Hamiltonian of this superconductor is given by
\be
H_{2S}=\int d^2x~ \chi(\v x)^T \left(-i\s_x\p_x-i\s_z\p_y+i\phi~i\s_y\right)\chi(\v x).\label{h2s}
\ee
Here $\phi>0$ and $\phi<0$ are inequivalent superconducting phases.
The subscript $ S $ stands for superconductor.  Tuning $\phi$ across zero induces a SPT phase transition. The critical point at $\phi=0$ is protected by an anti-unitary symmetry. The   mode space representation of the generator of such symmetry is $i\s_y$. Following Table \ref{tab:bulkbdd} we construct the Hamiltonian for a 3D bulk SPT so that its boundary theory is \Eq{h2s} with $\phi=0$:
\be
H_{3S}=\int d^3x~ X(\v x)^T \left(-i\s_x\t_z\p_x-i\s_z\t_z\p_y-i\s_0\t_x\p_z+i\lambda~i\s_0\t_y\right)X(\v x).\label{h3s}
\ee
The symmetry group of \Eq{h3s} is $\emptyset\times Z_2^T=Z_2^T$. The bulk symmetry $Z_2^T$ is generated by the mode space matrix $T=i\s_y\t_z$. The classification of SPTs in this symmetry class is $Z$, and \Eq{h3s} is the minimal model.
Again, the lattice version of \Eq{h3s} (see \ref{regr}) with $\lambda<0$ possesses gapless boundary modes (see \ref{reg}). These gapless boundary modes described by \Eq{h2s} at $\phi=0$, namely, 
the theory point of the 2D SPT phase transition.\\

In the next section we show that an interacting version of the holographic bulk 
can be viewed as a condensation of $Z_2^T$ domain walls. This is exactly analogous to the bosonic version of the holographic theory in Ref.\cite{Tsui2015}. Moreover, it turns out that this connection allows us to establish the holographic theory for interacting fermion SPT phase transitions.\\

\section{The interacting holographic bulk theory and condensed $Z_2^T$ domain walls} \label{ddw}
We start with the following interacting version of the holographic bulk theory
\be
H_{\rm Bulk,int}&&=\int d^{d+1}x\Big\{X^T(\v x)\Big[\p_t-i\sum_{j=1}^{d+1} \G_j\p_j+i \lambda M\Big]X(\v x)\nn&&-{u\over 2}\int d^{d+1} x \left[i X^T(\v x) m_0\otimes\tau_z X(\v x)\right]^2.\label{intD}
\ee 
Since the bulk is gapped, as long as the interaction term does not collapse the bulk gap (e.g. by considering a sufficiently weak $u$), it should not affect the topological properties of the bulk SPT.\\

\begin{figure}[h]
\centering
\includegraphics[width=0.5\linewidth]{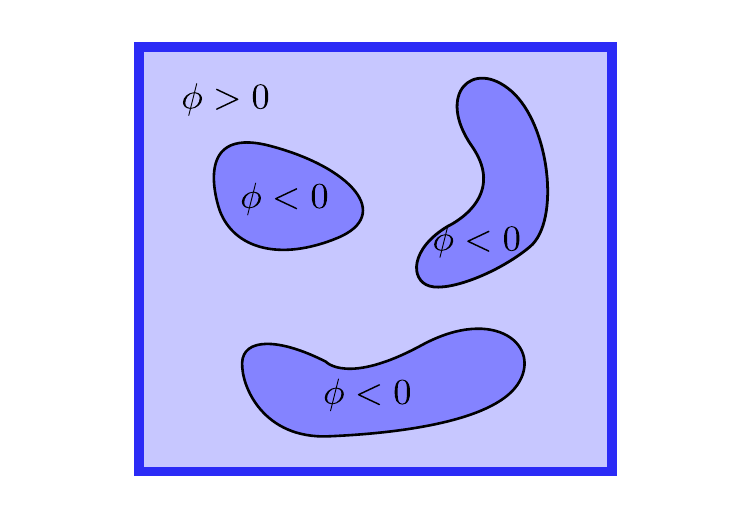}
\caption{(Color online) A snapshot of a fluctuating $ \phi(\v x, t) $ configuration described by the action in \Eq{hsD}. Here the dark blue regions have $ \phi<0 $ and the light blue regions have $ \phi >0 $. The illustration assumes $D=2$. }
\label{fig:snapshot}
\end{figure}

By Hubbard-Stratonavich decoupling the interaction term we can write the following path integral representation of \Eq{intD} as \be
Z&&=\int D[\phi(x,t)]~D[X(x,t)]~ e^{-S_{int}}\nn
S_{int}&&=\int d^{d+1}x~dt~\Big\{X^T(\v x,t)\Big[\p_t-i\sum_{j=1}^{d+1} \G_j\p_j+i\phi(\v x,t)m_0\otimes\tau_z+i\lambda M\Big]X(\v x,t)\nn&&+{1\over 2u} \phi(\v x,t)^2\Big\}\label{hsD}
\ee
According to Table \ref{tab:bulkbdd} $m_0\otimes\tau_z$ is the generator of the anti-unitary $Z_2^T$ symmetry. Consequently the term $i\phi(\v x,t)m_0\otimes\tau_z$ induces a dynamic breaking of $Z_2^T$. Because $\phi(\v x,t)$ fluctuate randomly, at any instant of time there are positive and negative spatial regions as shown in \Fig{fig:snapshot}. In other words the interacting bulk theory can be viewed as consisting of condensed $Z_2^T$ domain walls. Since for sufficiently weak $u$ the interaction term does not  affect the bulk SPT qualitatively, this proves that we can view the holographic bulk as consisting of condensed $Z_2^T$ domain walls.\\

\section{The decorated domain walls}\label{decdw}
In this section we show that the $Z_2^T$ domain walls discussed in the last section are  decorated with a topological non-trivial  $d$-dimensional SPT. Here we shall first present the numerics for the 1D and 2D examples supporting this claim. We leave the general analytical theory to \ref{apdx:ddw}.\\

For both the 1D (\Eq{hs}) and 2D (\Eq{hss2}) path integrals a snapshot of the $\phi$ configuration will have regions of positive $\phi$ surrounded by regions of negative 
$\phi$. The question is what happens on the domain walls.\\

To answer the above question we consider the following bulk Hamiltonian describing a frozen $\phi$ configuration in the holographic bulk SPT 

\be
&&H_{2,\phi}=\int d^2x~ \Psi^{\dagger}(\v x)[-i\s_z \t_z\p_x -i\s_0\t_x\p_y+\lambda\s_0\t_y + \phi(\v x)\s_x\t_z]\Psi(\v x)\label{eq:hdw}\\
&&H_{3S,\phi}=\int d^3x~ X(\v x)^T \Big[-i\s_x \t_z \p_x -i \s_z \t_z \p_y-i\s_0\t_x\p_z+i\lambda~i\s_0\t_y\nn&&+i\phi(\v x)~i\s_y\t_z\Big]X(\v x).\label{h3sp}
\ee
 Here $ |\phi(\v x)|\ll |\lambda|$ and is frozen in time. The regions where $ \phi(\v x) $ is positive/negative are $T$-breaking domains. It turns out that the domain walls, which have one lower dimension, are decorated with the non-trivial lower dimensional SPTs described by \Eq{eq:H1d} or \Eq{h2s}. In the following we present numerical results supporting this claim.\\ 

\begin{figure}[h]
\centering
\includegraphics[scale=1.5]{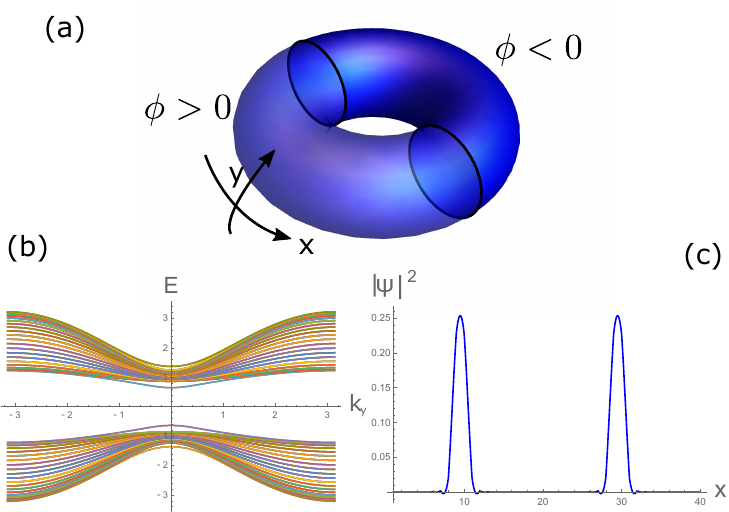}
\caption{(a) Geometry of the system studied in our numerical simulation in section \ref{sh}. We study a 2D system with periodic boundary conditions in $ x $ and $ y $. There are two domain walls running in the $ y $ direction. (b) Single body eigenvalue plot of \eqref{eq:hdlat}. There are two in-gap bands, each is two-fold degenerate. The parameters used are $ \lambda=-0.8 $, $ m_{\e}=0.3 $, $ n_x=40 $. (c) The sum of modulus square of the eigenfunctions of the in-gap bands at $ k_y=0 $. They are seen to be localized at the domain wall ($ x=9.5 $ and $ x=29.5 $).}
\label{fig:dw}
\end{figure}

\subsection{The domain wall in the 2D holographic bulk theory}\label{sh}
We study a lattice system with periodic boundary condition and two domain walls. The lattice model (presented in \Eq{eq:hdlat} in \ref{apdx:lat}) is constructed such that \Eq{eq:hdw} is the low energy effective theory. We then freeze the $\phi$ values such that there are two $\phi$ domain walls running parallel to $\hat{y}$ (see \Fig{fig:dw}(a)). The energy eigenvalues as are plotted a function of $ k_y $ in \Fig{fig:dw}(b).  We observe two degenerate in-gap bands each localized on a domain wall. These are the bands associated with the 1D SPT decorating each wall.  We verify this by plotting the sum of modulus square of the in-gap energy eigenfunctions at $k_y=0$ as a function of $ x $, and note that the result peaks at the locations of the domain wall.(See \Fig{fig:dw}(c).)
\\

\begin{figure}
\centering
\includegraphics[scale=1.3]{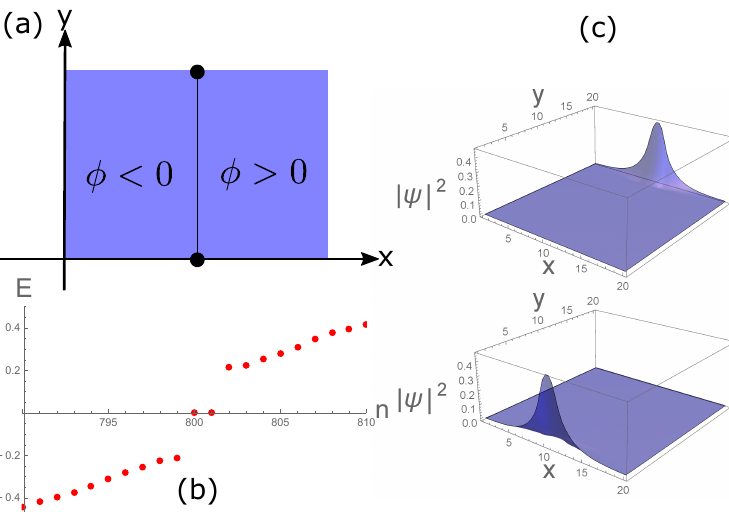}
\caption{(a) Geometry of the system studied in our numerical simulation in section \ref{sh}. We study a 2D system with open boundary conditions in $ x $ and $ y $. There is one domain wall running in the $ y $ direction. (b) Eigenvalue plot of \eqref{eq:hdlat}. $ n $ labels the eigenvalue number. There are two zero modes. The parameters used are $ n_x=n_y=20 $, $ \lambda=-0.8 $, $ m_{\e}=0.3 $. (c) The sum of modulus square of the eigenfunctions of the zero modes. Each eigenfunction is localized at a intersection of the domain wall and the boundary.} \label{fig:dwbdd}
\end{figure}

To illustrate that the 1D domain wall is decorated with the non-trivial SPT described by \Eq{eq:H1d}, we subject the lattice model to open boundary condition in both $x$ and $y$ directions and with a single frozen domain wall(see \Fig{fig:dwbdd}(a)) running in the $y$ direction. Diagonalization of the Hamiltonian on a finite lattice yields \Fig{fig:dwbdd}(b). There are two zero modes localized at the intersection of the domain wall and the boundary (\Fig{fig:dwbdd}(c)). These are the gapless modes at the end of the non-trivial 1D SPT. Since the 1D SPT is a topological insulator these are complex fermion zero modes.\\

\subsection{The domain wall in the 3D holographic bulk theory}\label{3dsc}
\begin{figure}
\centering
\includegraphics[width=1\linewidth]{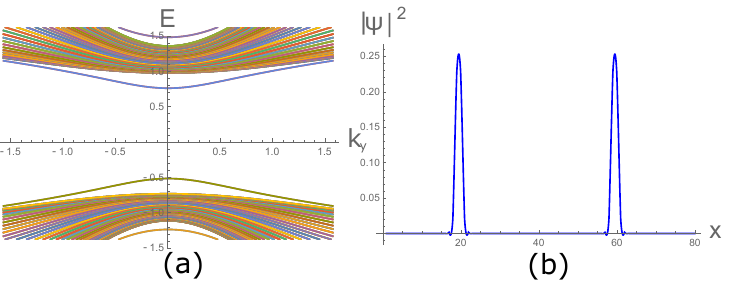}
\caption{(a) The single body eigenvalue plot of \eqref{eq:hdlat3d} discussed in section \ref{3dsc}. We study a 3D system with periodic boundary conditions in $ x $, $ y $ and $ z $. There are two in-gap bands, each is two-fold degenerate. The parameters used are $ \lambda=-0.8 $, $ m_{\e}=0.3 $, $ n_x=80 $ and $ k_z=0 $. (b) The sum of modulus square of the eigenfunctions of the in-gap bands at $ k_y=k_z=0 $. They are seen to be localized at the domain wall ($ x=19.5 $ and $ x=59.5 $).
}
\label{fig:dw3d}
\end{figure}

We first study a lattice under  periodic boundary condition with two $\phi$ domain walls running parallel to the $y$-$z$ planes. The lattice model presented in \ref{apdx:lat3d} is constructed such that \Eq{h3sp} is the low energy effective theory. The energy eigenvalues as are plotted for $k_y=0$ as a function of $ k_z $ in \Fig{fig:dw3d}(a).  We observe two degenerate in-gap bands, each localized on a domain wall. These are the bands associated with the 2D SPT decorating each wall. We verify this by plotting the sum of the modulus square of these in-gap energy eigenfunctions at $k_y=k_z=0$ as a function of $ x $, and note that it peaks near the locations of the domain walls. (See \Fig{fig:dw3d}(b).)\\

\begin{figure}
\centering
\includegraphics[scale=1.2]{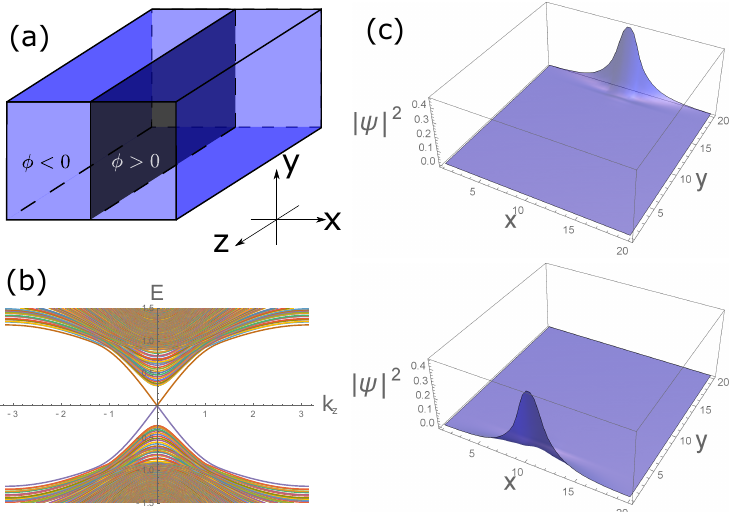}
\caption{(a) Geometry of the system studied in our numerical simulation in section \ref{3dsc}. The front and the back faces are meant to be identified. We study a 3D system with open boundary conditions in $ x $ and $ y $ but periodic boundary condition in $ z $. There is one domain wall parallel to the $ y $-$ z $ direction. (b) Eigenvalue plot of \eqref{eq:hdlat3d}. There are two non-degenerate gapless bands. The parameters used are $ n_x=n_y=20 $, $ \lambda=-0.8 $, $ m_{\e}=0.3 $. (c) The sum of modulus square of the eigenfunctions of the gapless bands at $ k_z=0 $. Each eigenfunction is localized at a intersection of the domain wall and the boundary.}
\label{fig:dwbdd3d}
\end{figure}

To illustrate that the 2D domain wall is itself a non-trivial 2D SPT (described by \Eq{h2s}) we subject the lattice model to open boundary conditions in both $x$ and $y$ but periodic in $z$. We freeze in a single domain wall running in the $y$-$z$ direction (see \Fig{fig:dwbdd3d}(a)).  Diagonalization of the Hamiltonian on finite lattice yields \Fig{fig:dwbdd3d}(b) which shows two non-degenerate gapless edge branches dispersing in the $z$ direction.  They are localized at the intersection of the domain wall and the boundary. This signifies the domain wall harbors a non-trivial 2D SPT. Since the 2D SPT is a chiral superconductor these edge modes are chiral Majorana in nature.\\


\section{The boundary of the interacting bulk theory}\label{bint}

Although the fluctuating scalar field in \Eq{hsD} does not affect the bulk properties, the same statement should not be made about the boundary hastily, because the latter is gapless. The boundary path integral is given by \be
Z&&=\int D[\phi(\v x,t)]~D[\chi(\v x,t)]~ e^{-S_{int}}\nn
S_{int}&&=\int d^dx~dt~\Big\{\chi^T(\v x,t)\Big[\p_t-i\sum_{j=1}^d \g_j\p_j+i\phi(\v x,t)m_0\Big]\chi(\v x,t)\nn&&+{1\over 2u} \phi(\v x,t)^2\Big\}.\label{hsd}
\ee
This path integral describes an interacting gapless fermion theory \be
H_{int}=H-{u\over 2}\int d^d x \left[i\chi^T(\v x) m_0\chi(\v x)\right]^2.\label{intd}
\ee
\\

 For weak $u$ the interaction term in \Eq{intd} can be viewed as a perturbation to the massless free fermion theory.  Simple dimension counting shows that for  $d>1$ the 
interaction is irrelevant at low energies. Thus for $d>1$ the boundary of the interacting bulk SPT (\Eq{intD}) is asymptotically described by the same massless free fermion theory.  However, for $d=1$ the interaction term is marginal, and a more careful consideration  is needed. We will do so in the 1D example below. In principle there could exist  a range of $u$ strong enough to cause spontaneous breaking of the $Z_2^T$ symmetry on the boundary but weak enough not to affect the bulk insulator properties\cite{DHLee,HongYao}.\\

\subsection{The interacting 1D boundary theory}
As argued in the last subsection the boundary of the interacting bulk theory is described by the following Hamiltonian
\be
H_{1,int}=H_1-{u\over 2}\int dx \left(\psi^\dagger(x) \s_x\psi(x)\right)^2,\label{int}
\ee
or by the following path integral
\be
Z_1&&=\int D[\phi(x,t)]D[\psi(x,t),\overline{\psi}(x,t)] e^{-S_{1,int}}\nn
S_{1,int}&&=\int dxdt \Big\{\overline{\psi}(x,t)\Big[\p_t-i\s_z\p_x+\phi(x,t)\s_x\Big]\psi(x,t)\nn&&+{1\over 2u} \phi(x,t)^2\Big\}.\label{hs}
\ee
\\ 




By bosonization we can map \Eq{int} to 
\be
H_{1,int}=\int dx\left\{{1\over 2}\left[\Pi(x)^2+\left(1+{u\over\pi}\right)(\p_x\varphi(x))^2\right]+{u\over 4\pi^2a^2}\cos\sqrt{16\pi}\varphi(x)\right\},
\label{sg1}\ee
where $\varphi$ is a scalar (real) boson field, $[\varphi(x),\Pi(y)]=i\delta(x-y)$, and $a$ is a short-distance cutoff. Equation(\ref{sg1}), the Sine-Gordon model, describes a gapless Luttinger liquid phase at $u<u_c$ where  $\<\psi^\dagger \s_x\psi\>\sim\<\sin\sqrt{4\pi}\varphi\>=0$. In this phase the $G\times Z_2^T$ symmetry is preserved. For $u>u_c$ the system enters a gapped phase with $\<\psi^\dagger \s_x\psi\>\ne 0$. In this phase the $Z_2^T$ symmetry is spontaneously broken. At $u_c$ a Kosterlitz-Thouless phase transition occurs.  The gapless Luttinger liquid is the interacting boundary theory of the bulk SPT. The modification of the boundary massless free fermion theory to a Luttinger liquid represents a non-perturbative effect of the interaction. \\


\subsection{The 2D example}   
The boundary of the interacting bulk theory is described by the following Hamiltonian
\be
H_{2S,int}=H_{2S}-{u\over 2}\int d^2x \left[\chi(\v x)^T \s_y \chi(\v x)\right]^2,\label{int2s}
\ee
or by the following path integral 
\be
Z_{2S}&=&\int D[\phi(x,t)]D[\chi(\v x,t)] e^{-S_{2S,int}}\nn
S_{2S,int}&=&\int d^2xdt \Big\{\chi(\v x,t)^T\left[\p_t-i\s_x\p_x-i\s_z\p_y+i\phi(\v x,t)~i\s_y\right]\chi(\v x,t)\nn&+&{1\over 2u} \phi(\v x,t)^2\Big\}\label{hss2} 
\ee
\\ 



Unlike the 1D case, the four fermion term in 
\Eq{int2s} is an irrelevant perturbation. Thus \Eq{int2s} describes an asymptotic  massless free fermion phase. Again, in principle there could exist  a range of $u$ strong enough to cause the spontaneous breaking of the $Z_2^T$ symmetry on the boundary but weak enough not to affect the bulk insulator properties.\\

\section{The phase transitions between some specific interacting SPTs}\label{phtrintspt}
So far we have discussed the free and interacting holographic bulk SPTs and their boundary theory. In this section we shall show that the interacting boundary theory serves as the critical theory for SPT transitions of interacting fermions. The only requirement we need to place on the strength of the interaction is that it does not close the energy gap of the bulk SPT.(We are also assuming the boundary SPTs in question remain inequivalent SPTs when interactions are allowed, and the bulk SPT remains non-trivial.) Under such conditions the interaction can either drive the boundary to spontaneously break the $Z_2^T$ symmetry, or it leaves $Z_2^T$ unbroken and the boundary remains gapless. In the latter case the boundary realizes the critical state of a continuous SPT transition. On the other hand, when the interaction causes the spontaneous breaking of the $Z_2^T$ symmetry, the boundary realizes the critical state of a first order SPT phase transition. 
In the following we discuss these interacting fermion SPT phase transitions for the 1D and 2D examples.

\subsection{The SPT transition between the interacting 1D SPTs}
\begin{figure}
\centering
\includegraphics[width=0.7\linewidth]{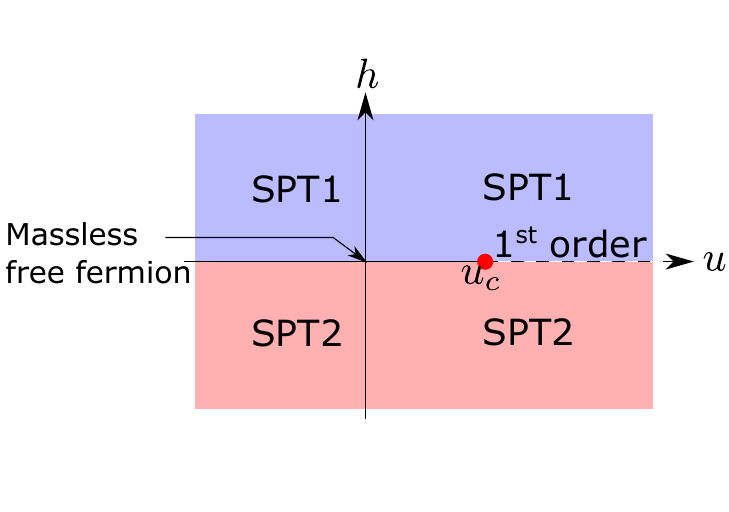}
\caption{(Color online) Schematic phase diagram of \eqref{int3}.  The line of $ h=0 $ preserves $ G\times Z_2^T $ and is protected from opening a gap. For $ u<u_c $ the system is in a gapless Luttinger liquid phase. For $ u>u_c $ the $ Z_2^T $ symmetry is spontaneously broken. The red dot marks the transition between the two, which is a Kosterlitz-Thouless phase transition. $ h $ is a relevent perturbation which opens a gap and leads to either the trivial or non-trivial SPT phase.}
\label{fig:phasediag}
\end{figure}

To study the SPT phase transition of the interacting 1D topological insulator consider the following Hamiltonian
\be
H_{1,int}^\prime=H_1-{u\over 2}\int dx \left(\psi^\dagger(x) \s_x\psi(x)\right)^2-h~\int dx~  \psi^\dagger(x) \s_x\psi(x),\label{int3}
\ee
where $h$ is the tuning parameter of the SPT transition.
The bosonized form of  \Eq{int3} is 
\be
H_{1,int}^\prime&=&\int dx \Big\{{1\over 2}\left[\Pi(x)^2+\left(1+{u\over\pi}\right)(\p_x\varphi(x))^2\right]\nn&+&{u\over 4\pi^2a^2}\cos\sqrt{16\pi}\varphi(x)\Big\}+{h\over\pi a}\int dx~\sin\sqrt{4\pi}\varphi(x).
\label{sg3}\ee
It can be shown that for $u>0$ the added last term is always relevant. Thus an infinitesimal $h$ induces an energy gap. The phases associated with opposite signs of $h$ correspond to inequivalent SPT phases. The gapless Luttinger liquid at $h=0$ is the critical state. For $u>u_c$ spontaneous symmetry breaking sets in. In that case tuning  $h$ from negative to positive induces a first order phase transition between the two SPT phases.  The phase diagram is shown in  \Fig{fig:phasediag}. The red dot marks the Kosterlitz-Thouless phase transition. It is a multi-critical point that requires fine tuning.

\subsection{The SPT transition between the interacting 2D SPTs}
To study the phase transition of the interacting 2D SPTs consider an Hamiltonian analogous to \Eq{int3}
\be
H_{2S,int}^\prime=H_{2S}-{u\over 2}\int d^2x \left[\chi(\v x)^T \s_y \chi(\v x)\right]^2-h\int d^2x \left[\chi(\v x)^T \s_y \chi(\v x)\right].\label{int2ss}
\ee
As discussed earlier at $h=0$ the interaction term is irrelevant hence the interacting massless fermion theory is asymptotically equivalent to a free massless theory. Under such condition the added $h$ term is a relevant perturbation and drives the system to gapped SPT phases. 
For sufficiently large $u$ spontaneous breaking of the $Z_2^T$ sets in at $h=0$. In that case tuning  $h$ from negative to positive induces a first order phase transition between the two SPT phases.  The phase diagram is similar to that in \Fig{fig:phasediag}. The only difference is that the red point no-longer describes the Kosterlitz-Thouless phase transition. Instead it is the multi-critical point marked by the spontaneous breaking of the $Z_2^T$ symmetry.\\

The SPT phase transitions under general conditions are qualitatively similar to those in the examples discussed above. Generically if the phase transition is continuous it is in the same universality class (except in 1D when the interaction is a marginal perturbation) as the free-fermion theory. Otherwise the phase transition is first order where the $Z_2^T$ symmetry is spontaneously broken at criticality. The only exception is the phase transition cutting across the multi-critical point (the red point in \Fig{fig:phasediag}). Here is the universality class can be very different from the free-fermion critical point. In  the literature a particularly simple case of this multi-critical point was studied where there is emergent super-symmetry.\cite{HongYao,Grover2014}   

\section{Bulk-Boundary Correspondence}\label{bbc}

The discussions in this paper and in Ref.\cite{Tsui2015} make the case that the critical point of the SPT phase transitions is the boundary theory of a fully gapped bulk SPT. In this section we ask how does this bulk-boundary correspondence help us understand the critical phenomena of the SPT phase transition. The answer is that the conformal spectrum of the critical theory is the entanglement spectrum associated with the ground state wavefunction of the holographic bulk SPT. 
The purpose of this section is to establish the above correspondence.
\\

For free fermion systems the correspondence of between the entanglement spectrum and the boundary spectrum has been established for ``spectral flattened'' Hamiltonian 
in Ref.\cite{Fidkowski2010}. Here by flattened Hamiltonian we mean the Hamiltonian that has the same eigenfunctions as the original Hamiltonian but its eigenvalues are flattened to $\pm 1$ depending on the sign of the original eigenvalues. By the holographic correspondence the boundary spectrum of the $ d+1 $ dimensional bulk SPT is the conformal spectrum of the $ d $-dimensional critical theory. Thus we have a simple example of the correspondence mentioned above. In this section we address problems that have emerging Lorentz invariance, but we do not require either non-interacting nor the spectrum flattening.\\

\begin{figure}
\centering
\includegraphics[width=1\linewidth]{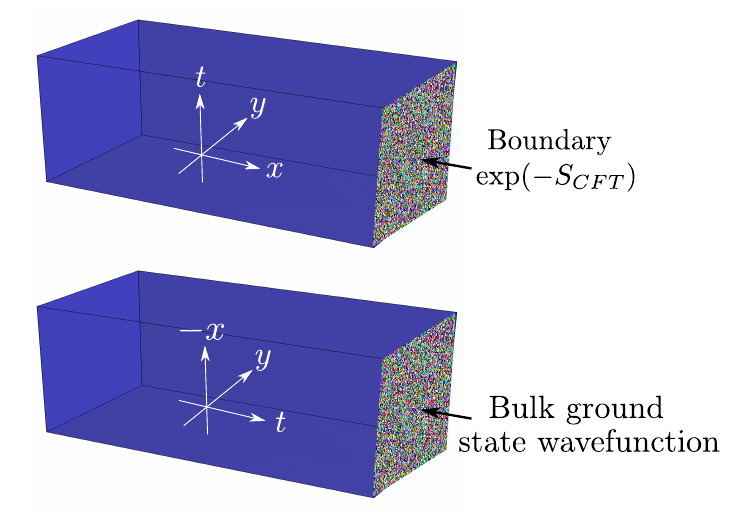}
\caption{(Color online) The upper figure shows a 2+1-D spacetime with an open boundary in the $ x $-direction. The boundary action, $ \exp(-S_{CFT}) $ is obtained by integrating over the bulk degrees of freedom in the bulk action. One may perform a space-time rotation in the $ x $-$ t $ plane, and regard the original $ x $-direction as the new time direction. This turns the original boundary action into the wavefunction of the groundstate wavefunction of a 2D Hamiltonian.}
\label{fig:bulkbdd}
\end{figure}

When the $ d+1 $-dimensional bulk (interacting) theory has Lorentz invariance, and is subjected to open boundary condition in, say, the $ x $-direction, one may perform a space-time rotation on the regularized lattice theory so that $x\ra t$ and $t\ra -x$.  After doing so the space-time of its boundary is rotated into pure space
 (see \Fig{fig:bulkbdd}). This predicts that the Boltzmann weight of the boundary theory, namely, the critical theory,  is equal to the ground state wavefunction of the bulk SPT (see \ref{apdx:spacetimelat}):
\begin{align*}
\exp(-S_{CFT}[\chi])=\Psi_{bulk}[\chi]
\end{align*}
If we perform a bi-partition cut perpendicular to the original time direction in the boundary wavefunction, the entanglement spectrum corresponds to the eigenvalues of the time direction transfer matrix in the boundary critical theory. The latter encodes the conformal spectrum. 
Hence we have a correspondence:
\begin{align*}
\text{Topological data} \lra \text{conformal data}.
\end{align*}
From a more conceptual point of view the above holographic correspondence implies that the topological classification of the holographic bulk also classifies the conformal field theory of the SPT transition.



\section{Acknowledgments}
This work was supported by the Theory Program at the Lawrence Berkeley National Laboratory, which is funded by the U.S. Department of Energy, Office of Science, Basic Energy Sciences, Materials Sciences and Engineering Division under Contract No. DE-AC02-05CH11231. L.M.T. acknowledges support from the Croucher Foundation's Fellowships for Postdoctoral Research.

\appendix
\section{The regularization rules}\label{regr}
In this section we start from the continuum  Hamiltonian in \Eq{eq:Hf},namely, 
\be
H=\int d^D x~ X^T(\v x) \left[\sum_{j=1}^{D}-i\Gamma_j\p_j +i \lambda M\right]X(\v x).\label{contin}
\ee
and try to regularize it on a lattice. 
The momentum space version of  \Eq{contin} is
\be
H=\sum_{\v k\in B(0)} X^T(-\v k) \left[\sum_{j=1}^{D}k_j\Gamma_j+i \lambda M\right]X(\v k).\label{contink}
\ee
Here $B(0)$ is a small ball center at $\v k=0$. The regularized Hamiltonian on a hyper-cubic lattice is given by
\be
H=\sum_{\v k\in {\rm BZ}} X^T(-\v k) \left[\sum_{j=1}^{D}\sin k_j\Gamma_j+i(D-\sum_{j=1}^{D}\cos k_j)M+ i \lambda M\right]X(\v k).\label{contink2}
\ee
Here ``BZ'' stands for the Brillouin zone of a $ D $-dimensional  hyper-cubic lattice. When $\lambda=0$ the second term removes the spurious gap nodes at all time-reversal invariant $\v k$ points except $\v k=0$. (A time-reversal invariant $\v k$ point satisfies $-\v k=\v k+\v G$ where $\v G$ is a reciprocal lattice vector.) Equation \ref{contink2} describes a non-trivial SPT when $\lambda<0$. \\
   
Fourier transform \Eq{contink2} back to the real space we obtain the following lattice Hamiltonian
\be
H&&=\sum_{\v x}\sum_{j=1}^D \left[X^T(\v x+\hat{e}_j)\left({-i\G_j-iM\over 2}\right)X(\v x)+ h.c.\right]\nn&&
+i(D+\lambda)\sum_{\v x} X^T(\v x)M X(\v x),\label{contink3} 
\ee
where $\hat{e}_j$ is the unit lattice vector in the $ j $th direction.  \\

Whether \Eq{contink2} describes a topologically non-trivial SPT depends on the sign of the ``low energy mass'' $\lambda$   relative to that of the ``regularization mass''. The sign of the regularization mass  is defined to be the (common) sign of $(D-\sum_{j=1}^D\cos k_j)$ at all $\v k\ne 0$ time-reversal invariant $\v k$ points. Clearly it is positive. When the sign of the low energy mass is opposite to that of the regularization mass, \Eq{contink2} and \Eq{contink3} describes a non-trivial SPT on a lattice. In \ref{reg} we show that the boundary of \Eq{contink3} is equivalent to the interface between two regions described by \Eq{contin} but with opposite $\lambda$. If the low energy mass has the same sign as that of the regularization mass, the boundary of \Eq{contink3} is equivalent to  the interface between two regions described by \Eq{contin} but with the same $\lambda$ sign. In that case there are no gapless boundary modes. Since the regularization mass is positive we conclude that when $\lambda<0$ \Eq{contink2} or \Eq{contink3} describes a non-trivial SPT.

\section{The unique mass matrix $m_0$ at $ n=n_0 $ there is only one} \label{apdx:m0}
The K-theory classification of free fermionic SPTs\cite{Schnyder2008,Kitaev2009,Wen2012} allows one to write down the form of allowed mass matrix $ M $ in \eqref{eq:Hf} consistent with the constraints. Here we cite the result. There are five cases where a SPT is non-trivial. In the following capital letters $ S$ an $A $ denote symmetric and anti-symmetric matrices respectively. Lower case letters  $s$ and $a$ denote symmetric and antisymmetric matrices which anti-commute with the gamma matrices of the minimal model. In addition, $I$ denotes identity matrix of appropriate dimension.

\begin{enumerate}

\item The mass matrix is written as $M=a_1\otimes S$ where $a_1^2=-I$. The requirement $M^2=-I$ implies $S^2=I$.

\item The mass matrix is written as $M=s_1\otimes A+a_1\otimes S$, where $s_1^2=-a_1^2=I$ and $\{s_1,a_1\}=0$. The requirement $M^2=-I$ implies $[A,S]=0$ and $S^2-A^2=I$. Hence $ S+A $ forms an orthogonal matrix.

\item The mass matrix is written as $M=s_1 \otimes A_1$, where $s_1^2=I$. The requirement $M^2=-I$ implies $A^2=-I$.

\item The mass matrix is written as $M=a_1\otimes S+s_1\otimes A_1+s_2\otimes A_2+s_3\otimes A_3$, where $-a_1^2=s_1^2=s_2^2=s_3^2=I$, $[a_1,s_i]=0$, $\{s_i,s_j\}=2\delta_{ij}$ and $ a_1 s_1s_2s_3 = -I $, 
which implies $ a_1 s_i = \frac{1}{2}\e^{ijk} s_j s_k $. The requirement $M^2=-I$ implies $S^2-A_1^2-A_2^2-A_3^2=I$ and $ \{S,A_i\} + \frac{1}{2}\e^{ijk}[A_j,A_k]=0 $. 
Hence $ H = S+iA_1+jA_2+kA_3 $ forms a Hermitian quaternion matrix $ H^{\dagger} = S^{T}-iA^{T}_1-jA^{T}_2-kA^{T}_3 = H $ which squares to $ I $.

\item The mass matrix is written as $M=s_1\otimes A+a_1\otimes S$, where $s_1^2=-a_1^2=I$ and $[s_1,a_1]=0$. The requirement $M^2=-I$ implies $\{A,S\}=0$ and $S^2-A^2=I$. Hence $ S+i A $ forms a hermitian matrix which squares to $I$.

\end{enumerate}

In all the above cases with the exception of case 3, there is only one symmetric matrix $ S $. At the smallest matrix dimension $ n=n_0 $, only a single 1-by-1 matrix $ S=1 $ is allowed. For case 3, at the smallest matrix dimension, only $ A = \e = i\s_y $ is allowed. Hence for all the non-trivial SPTs, there is only one mass matrix at $ n=n_0 $.

\section{The interface between two inequivalent minimal SPTs}\label{apdx:dwgapless}
In this section we show that in an SPT at $ n=n_0 $, on a domain wall where $ \phi $ changes sign, there exists gapless excitations.
Take \Eq{eq:Hf1} and let $ \phi(\v x) = \phi(x_1) $ be dependent on the first spatial coordinate only,
\begin{align}
H=\int d^d x~ \chi^T(\v x) \left[\sum_{j=1}^{d}-i\g_j\p_j +i \phi(x_1)~m_0\right]\chi(\v x).
\end{align}
where 
\begin{align*}
\phi(x_1) = 
\begin{cases}
+\phi_0~\text{if~}x_1>0\\
-\phi_0~\text{if~}x_1<0
\end{cases} 
\end{align*}
To find the one-body eigen-modes, we solve for the eigenvalue problem
\begin{align}
\left[\sum_{j=1}^{d}-i\g_j\p_j +i \phi(x_1)~m_0\right] \Phi (\v x) = E \Phi (\v x) \label{eq:ev}
\end{align}
since \Eq{eq:ev} is translational invariant in the $ x_2,x_3,...x_d $ directions, we can expand $ \Phi(\v x) $ in momentum eigenstates: $ \Phi(\v x) = \sum_{k_2,...,k_d} \Phi(x_1,\{k_j\}) e^{i\sum_{j=2}^dk_j x_j} $, thus \Eq{eq:ev} decouples into many independent 1D equations (one for each $ \{k_2,k_3,...,k_d\} $)
\begin{align}
\left[\sum_{j=2}^{d}\g_j k_j -i\g_1\p_1 +i \phi(x_1)~m_0\right] \Phi (x_1,\{k_j\}) = E \Phi (x_1,\{k_j\}). \label{eq:kev}
\end{align}
For $k_2=k_3=...=k_d=0$ \Eq{eq:kev} has a zero mode (\ie $ E=0 $) solution satisfying 
\begin{align*}
\left[ -i\g_1\p_1 +i \phi(x_1)~m_0\right] \Phi_0(x_1) = 0. \label{zm} 
\end{align*}
The solution of \Eq{zm} has the form
\begin{align*}
\Phi_0(x_1) =  \exp\left[\int_0^{x_1} \phi(x_1) \g_1 m_0\right] \Phi_0(0).
\end{align*}
A normalizable solution exists when $$ \g_1 m_0 \Phi_0(0)=-\Phi_0(0).$$

For non-zero $ \{k_2,...,k_d\} $, we can substitute $ \Phi(x_1,\{k_j\}) = \Phi_0(x_1) \widetilde{\psi}(\{k_j\}) $ into \Eq{eq:kev}, where $ \widetilde{\psi}(\{k_j\}) $ is a scalar function. The result is an eigenvalue equation for $ \widetilde{\psi}(\{k_j\})$:
\be
\left[\sum_{j=2}^{d}\g_j k_j \right] \widetilde{\psi}(\{k_j\}) = E \widetilde{\psi}(\{k_j\}).\label{zmk}
\ee
From \Eq{zmk} we see the one-body energy spectrum is gapless and is given by
\begin{align*}
E = \pm \sqrt{\sum_{j=2}^{d} k_j^2}.
\end{align*}
\section{The boundary of regularized minimal lattice models}\label{reg} 
As discussed in \ref{regr} the momentum space of a regularized SPT Hamiltonian on a hyper-cubic lattice is given by
$$
H=\sum_{\v k\in {\rm BZ}} X^T(-\v k) \left[\sum_{j=1}^{D}\sin k_j\Gamma_j+i(D-\sum_{j=1}^{D}\cos k_j)M+ i \lambda M\right]X(\v k).$$

In the presence of open boundary in, say, the $x_1$ direction and periodic boundary condition in $x_2,...,x_D$, we can partially Fourier transform the above equation w.r.t $x_1$ to obtain the following mixed real-momentum space Hamiltonian    
 \be
H&&=\sum_{\v q}\Big\{\sum_{x}\left[X^T(x+1,-\v q)\left({-i\G_1-iM\over 2}\right)X(x,\v q)+ h.c.\right]\nn&&
+\sum_{x}X^T(x,-\v q)\Big[\sum_{j=2}^{D}\sin k_j\Gamma_j+i(D-\sum_{j=2}^{D}\cos k_j)M+ i \lambda M\Big]X(x,\v q)\Big\}. \nn
\label{contink4} 
\ee
Here $\v q=(k_2,k_3,...,k_D)$. Because $\v q$ is a good quantum number \Eq{contink4} describes a collection of independent 1D chains, one for each $\v q$. \\

In particular for $\v q=0$ the eigen equation read
\be
&&\left({-i\G_1-iM\over 2}\right)\Phi(x+1,0)+\left({i\G_1-iM\over 2}\right)\Phi(x-1,0)+i(1+\lambda) M\Phi(x,0)\nn&&=E\Phi(x,0).
\ee
Multiply the above equation by $i\G_1$ we obtain
\be
&&\left({I+\G_1M\over 2}\right)\Phi(x+1,0)-\left({I-\G_1M\over 2}\right)\Phi(x-1,0)-(1+\lambda) \G_1M\Phi(x,0)\nn&&=iE\G_1\Phi(x,0).
\ee
This equation has a $E=0$ solution satisfying
\be
&&\left({I+\G_1M\over 2}\right)\Phi(x+1,0)-\left({I-\G_1M\over 2}\right)\Phi(x-1,0)\nn&&=(1+\lambda)\G_1M\Phi(x,0).
\ee

By opening up boundaries at $x=1$ and $x=L$, the $\Phi(x+1,0)$ or $\Phi(x-1,0)$ term should be neglected (set to be $0$) whenever the $x+1$ or $x-1$ is out of the region. We can solve the above equation by diagonalizing $\G_1M$. 
\\

In the $\G_1M=+1$ sector
 \be
\begin{cases}
\Phi_+(x+1,0)=(1+\lambda)\Phi_+(x,0) &\text{ for } x\neq L \\
 0 = (1+\lambda)\Phi_+(L,0) & \text{ for } x=L
\end{cases}
\ee And in the  $\G_1M=-1$ sector
 \be
\begin{cases}
\Phi_-(x-1,0)=(1+\lambda)\Phi_-(x,0) &\text{ for } x\neq 1 \\
 0 = (1+\lambda)\Phi_-(1,0) & \text{ for } x=1
\end{cases}
\ee
For $\lambda<0$ two normalizable solution exists, namely, 
\be 
&&\Phi_+(x)=(1-|\lambda|)^{x-1}\Phi_+(1) ~~\text{and}\nn
&&\Phi_-(x)=(1-|\lambda|)^{L-x}\Phi_-(L).\ee
The upper solution localizes on the left boundary, while the lower solution localizes on the right boundary. These are exactly the localized zero modes at the interface between $(\lambda>0,\lambda<0)$ and $(\lambda<0,\lambda>0)$, respectively since they live in the $ \G_1 M = +1 $ and $ \G_1 M = -1 $ sectors. Note that there is no localized zero mode for $\lambda>0$ due to the boundary constraint $0 = \Phi_+(L,0) $ and $0 = \Phi_-(1,0) $.

In the $\G_1M=+1$ sector
 \be
&&\Phi_+(x+1,0)=(1+\lambda)\Phi_+(x,0).
\ee And in the  $\G_1M=-1$ sector
\be
&&\Phi_-(x-1,0)=(1+\lambda)\Phi_-(x,0).
\ee
Let $x_1$ runs from $1$ to $L$, for $\lambda<0$ two normalizable solution exists, namely, 
\be 
&&\Phi_+(x)=(1-|\lambda|)^{x-1}\Phi_+(1) ~~\text{and}\nn
&&\Phi_-(x)=(1-|\lambda|)^{L-x}\Phi_-(L).\ee
The upper solution localizes on the left boundary, while the lower solution localizes on the right boundary. These are exactly the localized zero modes at the interface between $(\lambda<0,\lambda>0)$ and $(\lambda>0,\lambda<0)$, respectively.

\section{Summary of the classification result for SPT satisfying $T\rtimes Q\ltimes C$ symmetry and the minimal model matrix dimension}\label{classt}
In Table \ref{classft} we summarize the classification result of free fermion SPT protected by, maximally, $T\rtimes Q\ltimes C$ symmetries. Here  $C_\pm$ and $T_\pm$ implies $C^2=\pm 1$ and $T^2=\pm 1$. $ d $ is the spatial dimension. Each entry is a tuple representing $ (\text{classification group},n_0 ) $. Here $ 0 $ denotes the trivial group with only the identity element. Throughout the tables $\{T,Q\}=\{C,Q\}=0$.
\begin{table}[htp]
\caption{Classification table for SPT protected by $ T $ and $ C $ but without $ Q $ symmetry.}
\centerline{
    \begin{tabular}{|c|*{13}{c|}}
	\hline
	  &  & \multicolumn{2}{c|}{} & \multicolumn{2}{c|}{}  & \multicolumn{4}{c|}{$[T,C]=0$} & \multicolumn{4}{c|}{$\{T,C\}=0$}\\
	  &  & \multicolumn{2}{c|}{T only} & \multicolumn{2}{c|}{C only} & \multicolumn{2}{c|}{$T_+$} & \multicolumn{2}{c|}{$T_-$} & \multicolumn{2}{c|}{$T_+$} & \multicolumn{2}{c|}{$T_-$}\\
	 $d$ & No sym & $T_+$ & $T_-$ & $C_+$ & $C_-$ & $C_+$ & $C_-$ & $C_+$ & $C_-$ & $C_+$ & $C_-$ & $C_+$ & $C_-$\\
	\hline
	0 & $Z_2$,2 & $Z_2$,2 & 0,4 & $Z_2$,2 & $Z$,2 & $Z_2$,2 & 0,4 & 0,4 & 0,4 & $Z_2$,4 & $Z$,2 & $Z_2$,4 & $Z$,4\\
	\hline
	1 & $Z_2$,2 & $Z$,2 & $Z_2$,4 & $Z_2$,2 & 0,4 & $Z$,2 & $Z$,4 & $Z_2$,4 & $Z$,4 & $Z_2$,4 & 0,4 & $Z_2$,4 & 0,8\\
	\hline
	2 & $Z$,2 & 0,4 & $Z_2$,4 & $Z$,2 & $Z$,4 & 0,4 & 0,8 & $Z_2$,4 & 0,8 & $Z$,4 & 0,8 & $Z$,4 & $Z_2$,8\\
	\hline
	3 & 0,4 & 0,8 & $Z$,4 & 0,4 & 0,8 & 0,8 & $Z$,8 & $Z$,4 & $Z$,8 & 0,8 & 0,16 & 0,8 & $Z_2$,8\\
	\hline
	4 & 0,8 & 0,16 & 0,8 & 0,8 & $Z$,8 & 0,16 & 0,16 & 0,8 & 0,16 & 0,16 & $Z$,16 & 0,16 & $Z$,8\\
	\hline
    \end{tabular}
}
\caption{Classification table for SPT protected by $ T $ and $ C $ with $ Q $ symmetry.}
\centerline{
    \begin{tabular}{|c|*{13}{c|}}
	\hline
	  &  & \multicolumn{2}{c|}{} & \multicolumn{2}{c|}{}  & \multicolumn{4}{c|}{$[T,C]=0$} & \multicolumn{4}{c|}{$\{T,C\}=0$}\\
	  &  & \multicolumn{2}{c|}{T only} & \multicolumn{2}{c|}{C only} & \multicolumn{2}{c|}{$T_+$} & \multicolumn{2}{c|}{$T_-$} & \multicolumn{2}{c|}{$T_+$} & \multicolumn{2}{c|}{$T_-$}\\
	 $d$ & No sym & $T_+$ & $T_-$ & $C_+$ & $C_-$ & $C_+$ & $C_-$ & $C_+$ & $C_-$ & $C_+$ & $C_-$ & $C_+$ & $C_-$\\
	\hline
	0 & $Z$,2 & $Z$,2 & $Z$,4 & $Z_2$,4 & 0,4 & $Z_2$,4 & 0,4 & 0,8 & 0,8 & $Z_2$,4 & 0,4 & 0,8 & 0,8\\
	\hline
	1 & 0,4 & 0,4 & 0,8 & $Z_2$,4 & 0,8 & $Z$,4 & 0,8 & $Z_2$,8 & $Z$,8 & $Z$,4 & 0,8 & $Z_2$,8 & $Z$,8\\
	\hline
	2 & $Z$,4 & 0,8 & $Z_2$,8 & $Z$,4 & $Z$,8 & 0,8 & 0,16 & $Z_2$,8 & 0,16 & 0,8 & 0,16 & $Z_2$,8 & 0,16\\
	\hline
	3 & 0,8 & 0,16 & $Z_2$,8 & 0,8 & 0,16 & 0,16 & $Z$,16 & $Z$,8 & $Z_2$,16 & 0,16 & $Z$,16 & $Z$,8 & $Z_2$,16\\
	\hline
	4 & $Z$,8 & $Z$,16 & $Z$,8 & 0,16 & $Z_2$,16 & 0,32 & 0,32 & 0,16 & $Z_2$,16 & 0,32 & 0,32 & 0,16 & $Z_2$,16\\
	\hline
    \end{tabular}
}
\label{classft}
\end{table}
\section{The uniqueness of $ M $ in Table \ref{tab:bulkbdd}} \label{unique}
Here we present an argument that the proposed $ M=I_{n_0}\otimes i\t_y $ is the only mass matrix in the $ d+1 $-dimensional bulk consistent with the symmetries. We begin with the most general form of $ M $:
\begin{align*}
M=K_1\otimes \t_0+ K_2\otimes \t_x + K_3\otimes \t_z + S \otimes i\t_y 
\end{align*}
where $ K_i $ are antisymmetric and $ S $ is symmetric. To anti-commute with $ \Gamma_{d+1} $, $ K_1 $ and $ K_2 $ must be zero. So
\begin{align*}
M=K_3\otimes \t_z + S \otimes i\t_y 
\end{align*}
If $ K_3 $ were non-zero, then by commutation relationships of $ M $ with $ \Gamma_i $, $ U_{\alpha} $, $ A_{\beta} $ and the generator of $Z_2^T$, it can be seen that $ K_3 $ anticommutes with $ \gamma_i $, commutes with $ u_{\alpha} $, anticommutes with $ a_{\beta} $ but anticommutes with $ m_0 $. It means $ K_3 $ is a valid mass matrix in $ d $-dimensions not equal to $ m_0 $, contradicting the assumption that the boundary is at the minimal dimension $ n_0 $. So $ K_3=0 $. So we are left with
\begin{align*}
M= S \otimes i\t_y
\end{align*}
Similar analysis as before shows $ S $ commutes with $ \gamma_i $, $ u_{\alpha} $, $ a_{\beta} $ and $ m_0 $. Thus we can diagonalize $ S $ and $ H $ in \eqref{eq:Hf1} simultaneously. Also $ S^2=1 $ so its eigenvalues are $ \pm1 $. If $ S $ has both $ +1 $ and $ -1 $ sectors, then by projecting $ H $ to one of these sectors we would obtain a valid Hamiltonian with a smaller matrix dimension than $ n_0 $, which contradicts our assumption for $ n_0 $ being minimal. So $ S \propto I_{n_0} $ and $ M $ is the unique mass term consistent with all the symmetries.

\section{Deriving the bulk Hamiltonian (\Eq{eq:hbulk}) in section \ref{example1d} from the recipe of Table \ref{tab:bulkbdd}}\label{apdx:example}

In Majorana fermions, \Eq{eq:H1d} reads
\begin{align}
H_{1}=\int dx~\chi^T(x)\left[-i\rho_0\s_z\p_x+ \phi~ \rho_y \s_x\right] \chi(x)
\end{align}
where the $\rho_z=+1$ and $\rho_z=-1$ component of the Majorana fermion field are the real and imaginary parts of the complex  fermion field. There are two unitary symmetries. The charge $ U(1) $ symetry is generated by $ Q=i\rho_y\s_0 $ and the charge conjugation transformation is generated by $ C=\rho_z\s_z $.\\

Following the recipe of Table \ref{tab:bulkbdd}, we construct the following bulk (2D) Hamiltonian:
\begin{align}
H_{2}=\int d^2x~X^T(\v x)\left[-i\rho_0\s_z\t_z\p_x -i\rho_0\s_0\t_x \p_y + i\lambda~i \rho_0 \s_0 \t_y\right] X(\v x)
\end{align}
which is \Eq{eq:hbulk} in terms of Majorana fermions. According to the recipe, in the bulk, $ U(1) $ symmetry is generated by $ Q=i\rho_y\s_0\t_0 $, charge conjugation is generated by $ C=\rho_z\s_z\t_0 $, and the extra $ Z_2^T $ symmetry is generated by $ T= i\rho_y \s_x \t_z $. This would transform a complex creation operator into a complex annihilation operator. We may define another anti-unitary symmetry by combining the extra $ Z_2^T $ with the bulk charge conservation $ Q $ and charge conjugation $ C $, giving $ QCT=\rho_z i\s_y \t_z $. This operator maps $\psi \ra i\s_y\t_z \psi$ in the complex fermion language, which is the $ T $ symmetry in the bulk in section \ref{example1d}.\\


\section{Decorated domain wall interpretation of the bulk SPT} \label{apdx:ddw}
%
As discussed in the main text, the bulk Hamiltonian is given by

\begin{align*}
H_{d+1}=\int d^{d+1}x~ X^T(\v x)(-i\sum_{j=1}^{d+1}\Gamma_j \p_j+ i \lambda M) X(\v x)
\end{align*}

where
\be
&&\Gamma_j = \gamma_j \otimes \t_z ~\text{for $ j =1\dots d$}\\
&&\Gamma_{d+1} = I_{n_0} \otimes \t_x\\
&&M=I_{n_0} \otimes i\t_y
\ee
with unitary symmetries $ U_{\alpha}=u_{\alpha}\otimes \t_0 $, antiunitary symmetries $ A_{\beta}=a_{\beta}\otimes \t_z $, and an extra anti-unitary symmetry $ T=m_0\otimes \t_z $. 
Its boundary describes an SPT phase transition described by
\begin{align}
H_{d}=\int d^{d}x~ \chi^{T}(\v x)(-i\sum_{j=1}^{d}\gamma_j \p_j + i \phi m_0)\chi(\v x) \label{eq:Hbdd}
\end{align}

The boundary mass $ m_0 $ has the corresponding bulk term, $ M_{\e}=m_0\otimes \t_z $. This is so because by projecting $ M_{\e} $ into the boundary, we recover $ m_0 $. Also $ M_{\e} $ anti-commutes with all $ \G_j $'s and breaks the $Z_2^T$ symmetry in the bulk. In addition, $ M_{\e} $ also anti-commutes with $ M $. In the following we will study the domain walls of the $M_\e$ mass (i.e., the coefficient of $ M_{\e} $ changes sign) and show that they are decorated with the lower dimensional SPT.\\

Consider
\begin{align*}
H_{d+1}'=\int d^{d+1}x~X^{T}(\v x)(-i \sum_{j=1}^{d+1} \Gamma_j \p_j + i \e(x_1) M_{\e}+i \lambda M )X(\v x)
\end{align*}

Where $ \v x=(x_1,\dots,x_{d+1}) $, and $ \e(x_1) $ is a domain wall configuration in $ x_1 $ with 
\begin{align*}
\e(x_1) = 
\begin{cases}
+m_{\e}~\text{if~}x_1>0\\
-m_{\e}~\text{if~}x_1<0
\end{cases} 
\end{align*}

To find the one-body eigen-modes, we solve for the eigenvalue problem
\begin{align}
\left[-i \sum_{j=1}^{d+1} \Gamma_j \p_j + i \e(x_1) M_{\e}+i \lambda M\right] \Phi (\v x) = E \Phi (\v x)\label{eigmx}
\end{align}
Again we exploit the translational symmetry in $ x_2,x_3,...,x_{d+1} $ and go to the mixed real and momentum representation of $\Phi$, namely,  $ \Phi(\v x) = \sum_{\{k_j\}} \Phi(x_1,\{k_j\}) e^{i\sum_{j=2}^{d+1}k_j x_j}$. In the mixed representation 
\Eq{eigmx} becomes
\begin{align}
\left[\sum_{j=2}^{d+1} \Gamma_j k_j -i \G_1 \p_1 + i \e(x_1) M_{\e}+i \lambda M\right] \Phi(x_1,\{k_j\}) = E \Phi(x_1,\{k_j\}) \label{eq:ev2}
\end{align}

We first note that the $ x_1 $-dependent part of \eqref{eq:ev2} has a zero mode solution satisfying
\begin{align*}
\left[-i \G_1 \p_1 + i \e(x_1) M_{\e}\right] \Phi_0(x_1) = 0.
\end{align*}
The solution is 
\begin{align*}
\Phi_0(x_1) =  \exp\left[\int_0^{x_1} \e(x_1) \G_1 M_{\e}\right] \Phi_0(0).
\end{align*}
We see that in order for $ \Phi_0(x_1) $ to be normalizable, $\Phi_1(0)$ must satisfy $$ \G_1 M_{\e}\Phi_0(0) = -\Phi_0(0). $$  \\

The  solution of 
\Eq{eq:ev2} localized near $ x_1=0 $  is given by
 $$ \Phi(x_1,\{k_j\}) =  \exp\left[\int_0^{x_1} \e(x_1) \G_1 M_{\e}\right] \widetilde{\psi}(\{k_j\}) $$ where 
$$\G_1 M_{\e}\widetilde{\psi}(\{k_j\}) = -\widetilde{\psi}(\{k_j\}). $$
Note that $ \G_2,...,\G_{d+1} $, $ M $, $ U_{\a} $ and $ A_{\b} $ all commute with $ \G_1 M_{\e} $ and hence are block-diagonalized in the $-1$ eigenspace of $ \G_1 M_{\e} $ . After projecting to this eigenspace, \eqref{eq:ev2} becomes
\begin{align}
\left[\sum_{j=2}^{d+1} \G'_j k_j + i \lambda M'\right] \widetilde{\psi}'(\{k_j\}) = E \widetilde{\psi}'(\{k_j\}) \label{eq:ev2proj}
\end{align}
where the primed matrices/vectors are the projection of the original matrices/vectors. \Eq{eq:ev2proj} has the symmetries generated by the projected matrices $ U'_{\a} $, $ A'_{\b} $.\\

We note that \eqref{eq:ev2proj} has a gapped spectrum $ E=\pm\sqrt{\sum_{j=2}^{d+1}k_j^2 + \lambda^2}$. The solution is localized on the $M_\e$ domain wall hence corresponds to a $ d $-dimensional SPT protected by the same $G$ symmetry. 

\section{Real space lattice Models} \label{apdx:lat}
In this appendix we give the lattice models used for the numerical study in section \ref{decdw}.
\subsection{The 2D bulk}
In momentum space, the lattice model which recovers \Eq{eq:hdw} as the low energy theory is given by
\begin{align*}
H=\sum_{\v k} \Psi^{\dagger}(\v k)\left[\s_z\t_z\sin k_x+\s_0\t_x\sin k_y+(\lambda+2-\cos k_x-\cos k_y) M + \phi M_{\e}\right] \Psi(\v k),
\end{align*}
where $ M=\s_0 \t_y $, $ M_{\e}=\s_x \t_z $. Applying the regularization rules in \ref{regr} the lattice version of the above
equation is
\begin{align}
H={1\over 2}\sum_{\v x} &\left[ \Psi^\dagger(\v x)\left(-i\s_z\t_z-M\right)\Psi(\v x+\hat{x})+\Psi^\dagger(\v x)(-i\s_0\t_x-M)\Psi(\v x+\hat{y})+h.c.\right]\nn
&+ \Psi^\dagger(\v x)\left[ (\lambda+2) M + \phi(\v x) M_{\e}\right] \Psi(\v x) \label{eq:hdlat}
\end{align}
where $ \v x $ labels the lattice sites. Setting $ \phi(\v x)= m_{\e} sign(x-x_0) $ would fix a single domain wall at $ x=x_0 $.

\subsection{The 3D bulk} \label{apdx:lat3d}
In momentum space, the lattice model which recovers \Eq{h3sp} as the low energy theory is given by
\be
H&&=\sum_{\v k} X^{T}(-\v k)\Big[\s_x\t_z\sin k_x+\s_z\t_z\sin k_y+\s_0\t_x\sin k_z+i(\lambda+3-\cos k_x\nn&&-\cos k_y-\cos k_z) M + i\phi M_{\e}\Big] X(\v k)
\ee
where $ M=i\s_0 \t_y $, $ M_{\e}=i\s_y \t_z $. Applying the regularization rules in \ref{regr} we obtain the following lattice model
\be
&&H={1\over 2}\sum_{\v x} \Big[X^{T}(\v x)(-i\s_x\t_z-i M)X(\v x+\hat{x})+X^{T}(\v x)(-i\s_z\t_z-i M)X(\v x+\hat{y})\nn&&+X^{T}(\v x)(-i\s_0\t_x-i M)X(\v x+\hat{z})+h.c.\Big]
+ X^{T}(\v x)\Big[ i(\lambda+3) M+ i\phi(\v x) M_{\e}\Big] X(\v x)\nn \label{eq:hdlat3d}
\ee
where $ \v x $ labels the lattice sites. Setting $ \phi(\v x)= m_{\e} sign(x-x_0) $ would fix a single domain wall at $ x=x_0 $.

\section{Regularized Lattice theory on space time}\label{apdx:spacetimelat}
In this section we write down a regularized lattice space time model for \eqref{eq:Hf}. The continuum action is given by
\be
S = \int d^{D+1}x~X^T(t,\v x) [\p_0 + \sum_{j=1}^{D} -i \G_j \p_j + i \lambda M] X (t,\v x)+ S_{\rm int}[X(t,\v x)].
\label{sact}\ee

Going from space-time continuum to space-time lattice, we replace the time derivative term by a regularized lattice term:
\be
\int dx_0 X^T (t,\v x) \p_0  X(t,\v x) \ra \sum_{\omega} X^T(-\omega,\v x) \left[i\sin \omega + i (1-\cos \omega)M\right] X(\omega,\v x)
\label{tbp}
\ee
The corresponding term in the space-time lattice is given by
\be
\sum_t\left[ X^T(t+1,\v x)\left({I-i M\over 2}\right)X(t,\v x)+ h.c.\right]+i \sum_t X^T(t,\v x)M X(t,\v x).
\ee


On the other hand, the regularized free-fermion part of the Hamiltonian is achieved by the following replacement:
\begin{align}
&\int d^{D}x~X^T(t,\v x) \left[\sum_{j=1}^{D} -i \G_j \p_j + i \lambda M\right] X(t, \v x) \\
&\ra \sum_{\v k\in BZ}X^T(t,-\v k)\left\{\sum_{j=1}^{D} \left[\sin k_j\G_j + i(1-\cos k_j)M\right]+ i \lambda M\right\} X(t,\v k). \label{eq:Hbulk}
\end{align}
And as discussed in \ref{regr} the corresponding space-time lattice version is given by
\be
\sum_{\v x}\left[X^T(t,\v x+\hat{e}_j)\left({-i \G_j-i M\over 2}\right)X(t,\v x)+ h.c.\right]+i (D+\lambda) \sum_{\v x}X^T(t,\v x) M X(t,\v x).\nonumber
\ee

We assume the interaction part of the action is local in space-time and is Lorentz-invariant. In the following we shall determine transformed action after a space-time (Lorentz) transformation. Since the interaction part of the action is Lorentz-invariant we shall pay special attendion to the free fermion part in \Eq{tbp} and \Eq{eq:Hbulk}
\be
S_0&&= \sum_{\v p\in {\rm BZ}}X^T(-\v p) \Big\{ i \sin\omega + i (1-\cos \omega)M+ \sum_{j=1}^{D} \left[\sin k_j\G_j + i(1-\cos k_j)M\right]\nn&&+ i \lambda M\Big\} X(\v p) \nn
&&= \sum_{\v p\in {\rm BZ}}X^T(-\v p) M \Big\{ \sin\omega (-iM)+ i (1-\cos \omega) + \sum_{j=1}^{D} \Big[\sin k_j(-M\G_j)+ i(1-\cos k_j)\Big]\nn&&+i\lambda \Big\} X(\v p) \nn
&&=\sum_{\v p\in BZ}X^T(-\v p) M \Big\{\sin\omega \g_0+ i (1-\cos \omega) + \sum_{j=1}^{D} \Big[\sin k_j\g_j + i(1-\cos k_j)\Big]\nn&&+ i \lambda \Big\} X(\v p) \nn
&&=\sum_{\v p\in BZ}X^T(-\v p) M \Big\{\sin\omega \g_0+ i (1-\cos \omega) + \sum_{j=1}^{D} \Big[\sin k_j\g_j + i(1-\cos k_j)\Big]\nn&&+ i \lambda \Big\}X(\v p) \nn
&&=\sum_{\v p \in {\rm BZ}}X^T(-\v p)M\Big\{\sum_{\mu=0}^{D} \Big[\sin p_\mu\g_\mu + i(1-\cos p_\mu)\Big]+ i \lambda \Big\} X(\v p).
\label{long}\ee

In \Eq{long}  $\v p=(\omega,\v k)$ and ``BZ''stands for the space-time Brillouin zone. In addition, we defined $ \g_0 = -iM $ and $ \g_j = -M\G_j $. 
\\

Substitute $ X(\v p) = e^{i\frac{\pi}{4}\G_1} \widetilde{X}(\v p)$, $ X(-\v p)^T =  \widetilde{X}^T(-\v P)  e^{i\frac{\pi}{4}\G_1} $
\begin{align*}
S_0 &= \sum_{\v p\in {\rm BZ}}\widetilde{X}^T(-\v p)e^{i\frac{\pi}{4}\G_1}  M \Big\{ \sum_{\mu=0}^{D} \Big[\sin p_\mu\g_\mu + i(1-\cos p_\mu)\Big]+ i \lambda \Big\} e^{i\frac{\pi}{4}\G_1}\widetilde{X}(\v p)\\
&=\sum_{\v p\in {\rm BZ}}\widetilde{X}^T(-\v p) M e^{-i\frac{\pi}{4}\G_1}\Big\{ \sum_{\mu=0}^{D}\Big[\sin p_\mu\g_\mu + i(1-\cos p_\mu)\Big]+ i \lambda \Big\} e^{i\frac{\pi}{4}\G_1}\widetilde{X}(\v p) \\
&=\sum_{\v p\in {\rm BZ}}\widetilde{X}^T(-\v p) M\Big\{\sin\omega (-\g_1) + \sin k_1\g_0 + \sum_{j=2}^{D} \sin k_j\g_j \\
& ~~+ \sum_{\mu=0}^{D} i(1-\cos p_\mu)+ i \lambda \}\widetilde{X}(\v p)
\end{align*}

If we treat $ x_1 $ as the ``time" direction, the above action corresponds to a free fermion Hamiltonian
\be
H^\prime_0&&=\sum_{\v k^\prime\in {\rm BZ}^\prime}\widetilde{X}^T(-\v k^\prime)\Big\{\sin k^\prime_1 (-\G_1) + \sum_{j=2}^{D} \sin k^\prime_j\G_j
+ i\Big[\sum_{j=1}^D(1-\cos k^\prime_j)\nn&&+\lambda\Big]M \Big\}\widetilde{X}(\v k^\prime)
\ee
Here we have defined $\v k^\prime=(\omega,k_2,..,k_D)$, and BZ$^\prime$ stands for the Brillouin zone formed by $\v k^\prime$. \\

Due to the Lorentz invariance and the space-time local nature of the $S_{\rm int}$ the Lorentz-rotated interacting Hamiltonian is given by 
\be
\widetilde{H}=H^\prime_0+H_{\rm int}\label{rh}\ee
where $H_{\rm int}$ is the Hamiltonian correspond to $S_{\rm int}$. $\exp(-\e \widetilde{H})$ is the transfer matrix of the 
\Eq{sact} in the $ x $-direction. The Feynman amplitude between an initial and final field configuration after a long-``time'' propagation is the matrix elements of the projection operator to the the ground state wavefunction of $\widetilde{H}$. It is also the space-time Boltzmann weight of the gapless boundary theory. The preceding discussion corresponds to the following calculation:
\begin{align*}
&\exp(-S_{CFT}[\chi(x=T)])\exp(-S_{CFT}[\chi(x=-T)])^*\\
&=\int_{\chi'=\chi \text{ on boundaries}} \mathcal{D}[\chi'_{bulk}]\exp(-S_0[\{\chi'\}])\\
&=\bra{\{\chi(x=T)\}} \exp(-2T \widetilde{H}) \ket{\{\chi(x=-T)\}}\\
&=\exp(-2T E_0) \bra{\{\chi(x=T)\}}\psi_0\rangle \langle\psi_0\ket{\{\chi(x=-T)\}}\\
&\propto \Psi_{bulk}[\chi(x=T)] \Psi_{bulk}^{*}[\chi(x=-T)]
\end{align*}
where $ T\ra \infty $. $ \ket{\psi_0} $ and $ E_0 $ are the ground state wavefunction and energy, respectively. So we have 
\begin{align}
\Psi_{bulk}[\chi] = \exp(-S_{CFT}[\chi]) \label{eq:cftbulk}
\end{align} 
where we replaced the $ \propto $ sign by equality sign by assuming that a suitable constant has been added to $ S_{CFT} $ to normalize the RHS.

\subsection{Ground state entanglement spectrum = boundary conformal spectrum}
In this subsection we outline an argument for the equivalence between ground state entanglement spectrum and the boundary conformal spectrum, a generalization of \cite{Fidkowski2010} which proved the non-interacting fermion case. We illustrate our argument for $ 1+1 $-D boundary CFT/$ 2+1 $D bulk ground state but generalization to higher dimensions is straight-forward.

\begin{figure}[h]
\centering
\includegraphics[width=0.7\linewidth]{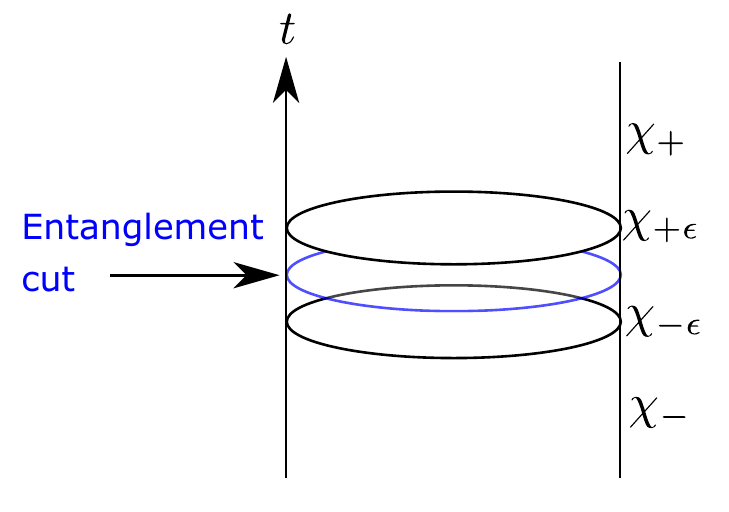}
\caption{(Color online) Illustration of infinite cylinder with entanglement cut perpendicular to $ t $ direction. The ground state living on the cylinder. The degrees of freedom are split into four regions $ \chi_{\pm} $, $ \chi_{\pm\e} $ as shown in figure. The blue circle is the bipartition cut.}
\label{fig:cyl}
\end{figure}

We study a ground state defined on a 2D infinite cylinder parameterized by $ (x,t) $, where the $ t $ direction is infinite. We assume the fields $ \chi $ are defined on discrete sites on the cylinder. Consider a bi-partition cut at $ t=0 $. We separate the fields into four regions: $ \chi_{+\e} $ denote the fields immediately above the cut(\ie $ t=\e>0 $), $ \chi_{-\e} $ denote the fields immediately below the cut(\ie $ t=-\e<0 $), $ \chi_{+} $ denote the fields above $ \chi_{+\e} $, (\ie $ t>\e $), and $ \chi_{-} $ denote the fields below $ \chi_{-\e} $, (\ie $ t<-\e $). See \Fig{fig:cyl} for illustration. Using \Eq{eq:cftbulk} we write the bulk ground state wave function as
\begin{align*}
\Psi_{bulk}[\chi_+,\chi_{+\e},\chi_{-\e},\chi_{-}] = \exp(-S_{CFT}[\chi_+,\chi_{+\e},\chi_{-\e},\chi_{-}])
\end{align*}
Since $ S_{CFT} $ is a local Lagrangian, we assume it can be split into three separate terms, each term involving only neighboring degrees of freedom
\begin{align*}
S_{CFT}[\chi_+,\chi_{+\e},\chi_{-\e},\chi_{-}] = S_{CFT}^+[\chi_+,\chi_{+\e}] + S_{CFT}^{\e}[\chi_{+\e},\chi_{-\e}] + S_{CFT}^-[\chi_{-\e},\chi_{-}]
\end{align*}

So
\begin{align*}
\Psi_{bulk}[\chi_+,\chi_{+\e},\chi_{-\e},\chi_{-}] = \phi^+[\chi_+,\chi_{+\e}]  \phi^{\e}[\chi_{+\e},\chi_{-\e}] \phi^-[\chi_{-\e},\chi_{-}]
\end{align*}
where we have defined $ \phi^{\pm/\e}[\chi] := \exp(-S_{CFT}^{\pm/\e}[\chi])  $. By trading terms amongst $ S_{CFT}^+,S_{CFT}^-,S_{CFT}^{\e} $, they can be defined to be suitably normalized.
\begin{align*}
\int \mathcal{D} \chi_+ |\phi^+[\chi_+,\chi_{+\e}]|^2 =1 \\
\int \mathcal{D} \chi_- |\phi^-[\chi_-,\chi_{-\e}]|^2 =1 \\
\int \mathcal{D} \chi_{+\e}\mathcal{D} \chi_{-\e} |\phi^{\e}[\chi_{+\e},\chi_{-\e}]|^2 =1
\end{align*}

The entanglement spectrum is defined to be the eigenvalues of
\begin{align*}
&\bra{\chi_+',\chi_{+\e}'} e^{-H_{ent}} \ket{\chi_+,\chi_{+\e}} = \int \mathcal{D}\chi_{-}\mathcal{D}\chi_{-\e}~ \Psi_{bulk}^*[\chi_+',\chi_{+\e}',\chi_{-\e},\chi_{-}] \Psi_{bulk}[\chi_+,\chi_{+\e},\chi_{-\e},\chi_{-}] \\
&=\int \mathcal{D}\chi_{-}\mathcal{D}\chi_{-\e}~ \phi^{+*}[\chi_+',\chi_{+\e}'] \phi^{\e*}[\chi_{+\e}',\chi_{-\e}] \phi^{-*}[\chi_{-\e},\chi_{-}] 
\phi^+[\chi_+,\chi_{+\e}] \phi^{\e}[\chi_{+\e},\chi_{-\e}] \phi^-[\chi_{-\e},\chi_{-}]\\
&=\phi^{+*}[\chi_+',\chi_{+\e}'] \phi^{+}[\chi_+,\chi_{+\e}]\int \mathcal{D}\chi_{-\e}~  \phi^{\e*}[\chi_{+\e}',\chi_{-\e}]\phi^{\e}[\chi_{+\e},\chi_{-\e}] 
\end{align*}

Note that
\begin{align*}
\phi^{\e}[\chi_{+\e},\chi_{-\e}] = \bra{\chi_{+\e}} e^{-2\e H_{CFT }} \ket{\chi_{-\e}}
\end{align*}
where $ H_{CFT} $ is the boundary Hamiltonian corresponding to $ S_{CFT} $.

So
\begin{align*}
&\bra{\chi_+',\chi_{+\e}'} e^{-H_{ent}} \ket{\chi_+,\chi_{+\e}} \\
&= \phi^{+*}[\chi_+',\chi_{+\e}'] \phi^{+}[\chi_+,\chi_{+\e}]\int \mathcal{D}\chi_{-\e}~  \bra{\chi_{+\e}} e^{-2\e H_{CFT }} \ket{\chi_{-\e}}\bra{\chi_{-\e}} e^{-2\e H_{CFT }} \ket{\chi_{+\e}'}\\
&= \phi^{+*}[\chi_+',\chi_{+\e}'] \phi^{+}[\chi_+,\chi_{+\e}] \bra{\chi_{+\e}} e^{-4\e H_{CFT }} \ket{\chi_{+\e}'}\\
e^{-H_{ent}} &= P^{\dagger} e^{-4\e H_{CFT }^T} P
\end{align*}
where $ \bra{\chi_{+\e}'} P \ket{\chi_+,\chi_{+\e}} = \delta_{\chi_{+\e}',\chi_{+\e}} \phi^{+}[\chi_+,\chi_{+\e}] $. It satisfies $ PP^{\dagger} = I $.

So for any eigenvector $ \Psi $ of $ e^{-4\e H_{CFT}} $ with eigenvalue $ e^{-4 \e E} $, $ P^{\dagger} \Psi $ is an eigenvector of $ e^{-H_{ent}} $ with the same eigenvalue. So entanglement spectrum contains the boundary CFT spectrum. Moreover the rank of matrix $ e^{-H_{ent}} $ equals that of $ e^{-4\e H_{CFT }^T} $. So its other eigenvalues are zero. So the ground state entanglement spectrum is equal to the boundary CFT conformal spectrum.

\newpage
\bibliographystyle{ieeetr}
\bibliography{bibs}

\end{document}